\documentclass{article}

\PassOptionsToPackage{numbers, sort&compress}{natbib}

\usepackage[main, final]{neurips_2026}     

\usepackage[utf8]{inputenc}
\usepackage[T1]{fontenc}

\usepackage{url}
\usepackage{booktabs}
\usepackage{amsfonts}
\usepackage{nicefrac}
\usepackage{microtype}
\usepackage{xcolor}

\usepackage{enumitem}
\usepackage{graphicx}
\usepackage{doi}
\usepackage{amsmath}
\usepackage{amsthm}
\usepackage{listings}
\usepackage{tcolorbox}
\tcbuselibrary{breakable}
\usepackage{algorithm}
\usepackage{algpseudocode}
\usepackage{placeins}
\usepackage{multirow}
\usepackage{pifont}
\usepackage{mdframed}
\usepackage{tabularx}

\newtheorem{theorem}{Theorem}

\newtheorem{assumption}{Assumption}

\theoremstyle{definition}
\newtheorem{definition}[theorem]{Definition}

\theoremstyle{remark}

\definecolor{promptbg}{RGB}{250,250,250}
\definecolor{promptframe}{RGB}{180,180,180}
\definecolor{titlebg}{RGB}{230,230,230}

\mdfdefinestyle{promptbox}{%
    backgroundcolor=promptbg,
    linecolor=promptframe,
    linewidth=0.8pt,
    roundcorner=2pt,
    innertopmargin=6pt,
    innerbottommargin=6pt,
    innerleftmargin=8pt,
    innerrightmargin=8pt,
    frametitlebackgroundcolor=titlebg,
    frametitlerule=true,
    frametitlerulewidth=0.4pt,
    frametitlefont=\small\bfseries\sffamily,
    frametitleaboveskip=3pt,
    frametitlebelowskip=3pt
}

\lstdefinestyle{prompt}{%
    basicstyle=\ttfamily\scriptsize,
    breaklines=true,
    breakatwhitespace=false,
    columns=flexible,
    keepspaces=true,
    showstringspaces=false,
    tabsize=2,
    aboveskip=0pt,
    belowskip=0pt
}

\definecolor{codegreen}{rgb}{0,0.6,0}
\definecolor{codegray}{rgb}{0.5,0.5,0.5}
\definecolor{codepurple}{rgb}{0.58,0,0.82}
\definecolor{backcolour}{rgb}{0.95,0.95,0.92}

\lstdefinestyle{mystyle}{
    backgroundcolor=\color{backcolour},
    commentstyle=\color{codegreen},
    keywordstyle=\color{magenta},
    numberstyle=\tiny\color{codegray},
    stringstyle=\color{codepurple},
    basicstyle=\ttfamily\footnotesize,
    breakatwhitespace=false,
    breaklines=true,
    captionpos=b,
    keepspaces=true,
    numbers=left,
    numbersep=5pt,
    showspaces=false,
    showstringspaces=false,
    showtabs=false,
    tabsize=2
}
\lstdefinelanguage{json}{
    basicstyle=\ttfamily\footnotesize,
    showstringspaces=false,
    breaklines=true,
    frame=lines,
    stringstyle=\color{codepurple},
    morestring=[b]",
    morecomment=[l]{//},
    morecomment=[s]{/*}{*/},
    morekeywords={true,false,null},
    sensitive=false
}

\hypersetup{
    pdftitle={ReLoop: Structured Modeling and Behavioral Verification for Reliable LLM-Based Optimization},
    pdfauthor={Anonymous},
    pdfkeywords={Automated Optimization Modeling, Behavioral Verification, Silent Failures, Large Language Models},
    colorlinks=true,
    linkcolor=blue,
    citecolor=blue,
    urlcolor=blue
}

\title{ReLoop: Structured Modeling and Behavioral Verification for Reliable LLM-Based Optimization}

\author{
  Junbo Jacob Lian$^{1,2,3,*}$\quad Yujun Sun$^{2,*}$\quad Huiling Chen$^{4}$ \\
  \textbf{Chaoyu Zhang$^{5}$\quad Hanzhang Qin$^{1,\dagger}$\quad Chung-Piaw Teo$^{1,\dagger}$} \\
  $^{1}$\,Institute of Operations Research and Analytics, National University of Singapore \\
  $^{2}$\,McCormick School of Engineering, Northwestern University \\
  $^{3}$\,Wenzhou Buyi Pharmacy Chain Co., Ltd. \\
  $^{4}$\,College of Computer Science and Artificial Intelligence, Wenzhou University \\
  $^{5}$\,Department of Decision Analytics and Operations, City University of Hong Kong \\
  \texttt{\{jacoblian,\,yujunsun2026\}@u.northwestern.edu},\ \ \texttt{chenhuiling.jlu@gmail.com} \\
  \texttt{cy.zhang@cityu.edu.hk},\ \ \texttt{\{hzqin,\,bizteocp\}@nus.edu.sg}
}

\begin{document}
\maketitle
{\renewcommand{\thefootnote}{}\footnotetext{$^{*}$Co-first authors.\quad $^{\dagger}$Corresponding author.}}
\begin{abstract}
Large language models (LLMs) can translate natural-language problem descriptions into optimization code, but the code is prone to \emph{silent failures}: it executes and returns a solver-feasible solution while encoding a semantically incorrect formulation. On compositional problems, the resulting feasibility--correctness gap reaches 90 percentage points.
We introduce \textbf{ReLoop}, which combines two mechanisms.
\emph{Structured generation} decomposes code production into a four-stage reasoning chain (understand, formalize, synthesize, verify) to reduce formulation errors during generation.
\emph{Behavioral verification} detects the errors that remain by testing whether the formulation responds correctly to solver-based parameter perturbation, a signal that comes from the solver rather than from LLM self-review and requires no ground truth.
The two mechanisms address different error structures: structured generation gives the largest gain on compositional problems (\textbf{+8.5pp} accuracy on RetailOpt-190 with Claude Opus~4.6), and behavioral verification gives its largest gain on localized defects (\textbf{+4.4pp} on MAMO-ComplexLP).
With diagnostic execution recovery, ReLoop reaches \textbf{100\% executable code} on Claude Opus~4.6, and relative to direct generation it raises or preserves every reported metric of the three chat-tuned foundation models on all three benchmarks.
For the narrowly fine-tuned SFT model we test, the chain-of-thought prompt conflicts with its learned output format and lowers its accuracy on MAMO-ComplexLP; we document and analyze this interaction.
We release \textbf{RetailOpt-190}, 190 compositional retail optimization scenarios in which several constraints interact.
Code and benchmark: \url{https://github.com/junbolian/ReLoop}.
\end{abstract}

\section{Introduction}

Optimization solvers cannot tell a correct model from an incorrect one that happens to be feasible.
For LLM-based optimization, this produces \emph{silent failures}: code that executes without errors and returns solver-feasible solutions, yet encodes a semantically incorrect formulation.
An inventory model that ignores perishability generates plans that hold expired stock, and a routing model that omits time-window constraints dispatches trucks that arrive after customers leave.
In both cases the solver reports an ``optimal'' solution to the wrong problem.

Silent failures are frequent on compositional problems, in which several constraints interact (capacity, perishability, substitution, lead times). On such problems, DeepSeek-V3.2 reaches 91.1\% solver feasibility but only 0.5\% formulation correctness, a 90-point \emph{feasibility--correctness gap} (\S\ref{sec:main_results}).
Existing verification does not detect these failures.
Solver feedback catches syntax errors but not missing constraints, and LLM self-critique~\citep{madaan2023selfrefine, huang2024selfcorrection} inherits the reasoning gaps that caused the errors.
Environment-guided refinement~\citep{shinn2023reflexion} uses execution feedback but cannot distinguish a feasible but wrong model from a correct one, and execution-based reranking~\citep{ni2023lever} requires ground-truth solutions, which are unavailable for a new problem.

Silent failures arise at the \emph{modeling} stage, when the LLM translates the natural-language problem into a mathematical formulation, and they persist because these post-hoc checks cannot detect them.
We therefore address them at two levels.

\emph{At the generation level}, an expert operations research modeler does not write code directly: the modeler decomposes the problem, writes the mathematical model with an explicit type decision for each variable, and then implements and cross-checks it.
We encode this workflow as a four-stage chain-of-thought prompt (understand, formalize, synthesize, verify), which reduces formulation errors before any external verification runs.

\emph{At the verification level}, a correct optimization model satisfies behavioral invariants that can be tested without ground truth.
(i)~\emph{Constraint presence}: shrinking a capacity limit to near zero should change the optimal objective, so no change indicates that the constraint is likely missing.
(ii)~\emph{Objective completeness}: perturbing a cost coefficient should shift the optimal objective, so no change indicates that the term is likely omitted.
Declarative self-review inherits the errors of the generating LLM~\citep{huang2024selfcorrection}, whereas solver-based perturbation provides an \emph{external semantic signal}, because the response is computed by the solver, not judged by the LLM.

\textbf{ReLoop} implements both levels with three components:
(1)~\emph{structured generation} with the four-stage prompt (\S\ref{sec:sg});
(2)~\emph{two-layer behavioral verification}, consisting of execution recovery with IIS-based diagnostics~(L1) and solver-based perturbation testing~(L2); and
(3)~\emph{diagnosis-guided repair} with a regression guard.

\textbf{Contributions.}
\begin{enumerate}[leftmargin=*, itemsep=2pt, topsep=3pt]
    \item We propose \emph{structured generation} for optimization code, which decomposes modeling into four stages with explicit variable-type reasoning and a grounded completeness check. On compositional problems, structured reasoning is the largest source of accuracy gains for the foundation models; we attribute this gain to explicit structured reasoning in general rather than to the OR-specific staging (\S\ref{sec:sg}, \S\ref{sec:ablation}).

    \item We formalize \emph{behavioral verification}, which tests whether a formulation responds correctly to solver-based parameter perturbation. To our knowledge, it is the first approach that uses solver-response perturbation as an external, non-LLM semantic signal for detecting silent failures without ground truth. It is the largest single source of accuracy gains on problems with localized defects (\S\ref{sec:verification}, \S\ref{sec:ablation}).

    \item We design \textbf{ReLoop}, a pipeline that combines structured generation, execution recovery, and behavioral testing with regression-guarded repair, and we release \textbf{RetailOpt-190}, 190 compositional retail optimization scenarios (\S\ref{sec:verification}, \S\ref{sec:benchmark}).

    \item A cross-benchmark ablation shows that structured generation and behavioral verification are \emph{complementary}, since each gives the larger gain under a different error structure, and that CoT can conflict with domain-specific fine-tuning (\S\ref{sec:experiments}).
\end{enumerate}

\section{Related Work}

\paragraph{LLM-Based Optimization Modeling.}
Methods that translate natural language into mathematical programs use prompting~\citep{ahmaditeshnizi2024optimus, xiao2024chainofexperts}, supervised fine-tuning~\citep{huang2025orlm, liu2024llmopt, lu2025optmath}, reinforcement learning with solver feedback~\citep{chen2025solverinformedrl}, and agentic workflows~\citep{liang2026leallmopt}.
All of them improve \emph{generation quality}, through training or prompting, and are complementary to ReLoop.
ReLoop contributes a domain-specific structured generation strategy, which is prompt-based and applies to any model, and adds post-generation \emph{behavioral verification}, which detects when a feasible model is semantically incorrect.

\paragraph{Verification and Self-Correction.}
Table~\ref{tab:related} groups existing verification methods by paradigm, and each paradigm has a limitation for optimization.
\emph{Solver-based feedback}~\citep{chen2025solverinformedrl, chen2025optichat} cannot detect a model that is feasible but wrong, which is exactly a silent failure.
\emph{LLM self-critique}~\citep{madaan2023selfrefine} fails because, without external feedback, a model cannot identify errors caused by its own reasoning gaps~\citep{huang2024selfcorrection}, and these are the gaps that produced the silent failure.
\emph{Environment-guided refinement}~\citep{shinn2023reflexion} incorporates execution feedback, but the solver status and the objective value do not distinguish a correct model from a feasible but wrong one.
\emph{Execution-based reranking} relies on training-time labels~\citep{ni2023lever} or generated test cases~\citep{chen2022codet}; in optimization, ground-truth programs are scarce, and generating reliable test cases requires the semantic understanding that the LLM lacks.
\emph{Semantic-anchor alignment} (SAC-Opt)~\citep{zhang2025sacopt} reconstructs semantic anchors from the generated code, aligns them with anchors extracted from the problem, and applies corrections where the two sets disagree; this backward-guided check operates in the LLM's semantic space. Because the LLM reconstructs both anchor sets, their agreement can depend on the same misreading that caused an error, which is the correlated-error mechanism identified for self-review~\citep{huang2024selfcorrection}. The two approaches are complementary: SAC-Opt compares semantic representations produced by the LLM, whereas ReLoop tests solver behavior.
ReLoop uses solver-based perturbation as an external semantic signal, which provides the independent feedback that \citet{huang2024selfcorrection} identify as necessary for reliable correction.

\paragraph{Sensitivity Analysis and Program Verification.}
Sensitivity analysis studies how the optimum changes with the parameters~\citep{bertsimas1997introduction, vanderbei2020linear}.
We use it for verification: instead of \emph{interpreting} sensitivity, we test whether it \emph{exists}, because a parameter that should affect the optimum but produces zero sensitivity indicates a missing component.
Testing expected sensitivity patterns is an instance of metamorphic testing~\citep{chen1998metamorphic, segura2016survey, chen2018metamorphic}: Assumption~\ref{prop:perturb} defines metamorphic relations specific to optimization, with the solver as the test oracle.
Classical program verification requires formal specifications~\citep{cousot1977abstract, clarke1999model}; our approach does not, because it tests properties that any correct optimization model satisfies.

\paragraph{Benchmarks.}
Existing benchmarks, including NL4Opt~\citep{ramamonjison2023nl4opt}, MAMO~\citep{huang2024mamo}, IndustryOR~\citep{huang2025orlm}, and OptMATH~\citep{lu2025optmath}, mostly test constraints in isolation: they evaluate whether a model can encode a single constraint type rather than several interacting ones.
RetailOpt-190 instead composes 8 scenario families in which multiple constraints must bind jointly (\S\ref{sec:benchmark}).

\begin{table}[t]
\centering
\small
\caption{Verification approaches for LLM-generated optimization. $^\dagger$Requires ground-truth labels.}
\label{tab:related}
\vspace{0.5em}
\begin{tabular}{@{}lcccc@{}}
\toprule
\textbf{Method} & \textbf{Detects Silent} & \textbf{External} & \textbf{No Ground} & \textbf{Iterative} \\
                & \textbf{Failures} & \textbf{Signal} & \textbf{Truth} & \textbf{Repair} \\
\midrule
Solver feedback~\citep{chen2025solverinformedrl} & \ding{55} & \ding{51} & \ding{51} & \ding{55} \\
OptiChat~\citep{chen2025optichat} & \ding{55} & \ding{51} & \ding{51} & \ding{51} \\
Self-Refine~\citep{madaan2023selfrefine} & \ding{55} & \ding{55} & \ding{51} & \ding{51} \\
Reflexion~\citep{shinn2023reflexion} & \ding{55} & \ding{51} & \ding{51} & \ding{51} \\
SAC-Opt~\citep{zhang2025sacopt} & \ding{51} & \ding{55} & \ding{51} & \ding{51} \\
LEVER$^\dagger$~\citep{ni2023lever} & \ding{51} & \ding{51} & \ding{55} & \ding{55} \\
\midrule
\textbf{ReLoop (ours)} & \ding{51} & \ding{51} & \ding{51} & \ding{51} \\
\bottomrule
\end{tabular}
\vspace{-1em}
\end{table}

\FloatBarrier

\section{Method}
\label{sec:method}

\begin{figure*}[t]
\centering
\includegraphics[width=0.9\textwidth]{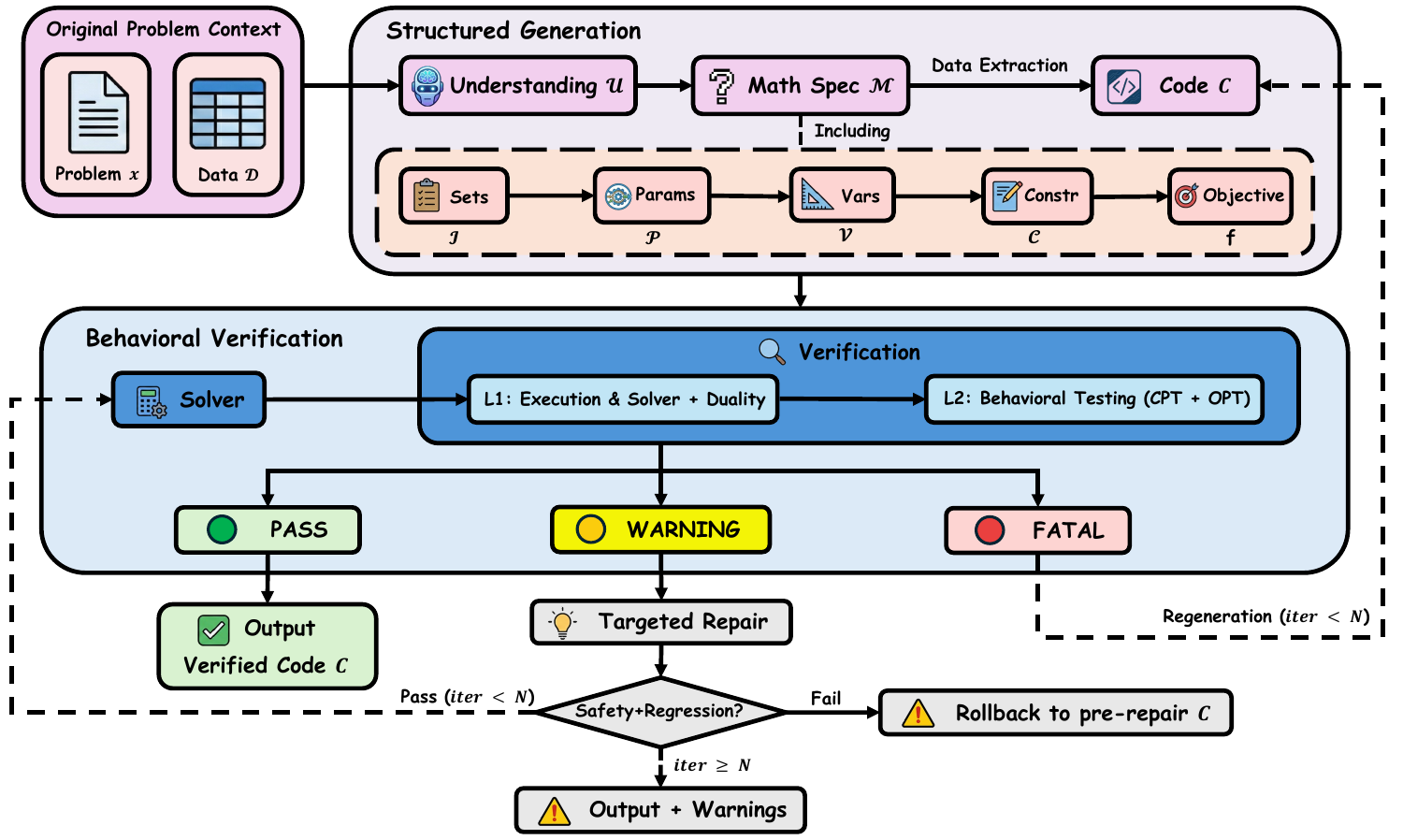} 
\caption{ReLoop overview. \textbf{Structured Generation} follows expert modeling practice: understand the problem, formalize the mathematical model with explicit variable-type reasoning, synthesize Gurobi code with data extraction, and self-verify completeness. \textbf{Behavioral Verification}: L1 checks execution correctness (\textsc{Fatal} blocks output); L2 tests constraint (CPT) and objective (OPT) presence via solver-based perturbation (\textsc{Warning}/\textsc{Pass}). \textbf{Diagnosis-Guided Repair}: \textsc{Fatal} triggers diagnosis-specific regeneration (syntax tracebacks, IIS, or unbounded rays); \textsc{Warning} triggers targeted repair with regression rollback. When the budget~$N$ is exhausted, ReLoop returns the best verified code.}
\label{fig:framework}
\end{figure*}

ReLoop combines two mechanisms. \emph{Structured generation} (\S\ref{sec:sg}) decomposes code production into a four-stage reasoning chain that follows expert modeling practice, which reduces formulation errors. \emph{Behavioral verification} (\S\ref{sec:verification}) detects the errors that remain by testing how the formulation responds to solver-based perturbation. Execution recovery (L1) and diagnosis-guided repair (\S\ref{sec:dr}) complete the pipeline. Figure~\ref{fig:framework} shows the architecture, and Algorithm~\ref{alg:reloop} (Appendix~\ref{app:reloop_details}) states it formally.

\subsection{Problem Statement}
\label{sec:problem}

Let $x$ denote a natural-language optimization problem description containing all parameter values.

\begin{definition}[Semantic Correctness]
\label{def:semantic}
Code $C$ is \emph{semantically correct} for $x$ if the optimization model encoded by $C$ captures all constraints and objective terms specified in $x$, i.e., the feasible region and objective function match the intended formulation.
\end{definition}

\begin{definition}[Silent Failure]
\label{def:silent}
Code $C$ is a \emph{silent failure} if it (i)~executes without error, (ii)~returns a solver-feasible solution, yet (iii)~is not semantically correct.
\end{definition}

Standard software testing checks execution success and solution feasibility, so it cannot detect silent failures. Our verification approach uses a behavioral invariant of correct optimization models:

\begin{assumption}[Perturbation Sensitivity]
\label{prop:perturb}
Let $z^*(\theta)$ denote the optimal objective of a correctly formulated model parameterized by $\theta$, and let the response ratio to a perturbation $\theta_i \to \tilde{\theta}_i$ be $r_i = |z^*(\tilde{\theta}_i) - z^*(\theta)|\,/\,|z^*(\theta)|$. Assume $\theta_i$ governs a component that (i)~binds, or can be forced to bind under sufficiently extreme perturbation, and (ii)~carries a non-negligible objective contribution. Then $r_i \gg 0$ under extreme perturbation. This is a design assumption rather than a theorem; the graduated thresholds of \S\ref{sec:l2} operationalize it, and two guards in the main pipeline handle its edge cases: when $z^*(\theta) = 0$ the implementation falls back to the absolute change $|z^*(\tilde{\theta}_i) - z^*(\theta)|$, and perturbation-induced infeasibility is treated as a strong \textsc{Pass} signal (Appendix~\ref{app:reloop_details}).
\end{assumption}

\noindent The contrapositive gives our detection criterion: if perturbing a parameter that \emph{should} affect the objective produces a negligible change, the corresponding model component is likely absent from the generated code. Extreme perturbation can also force natively inactive constraints to bind (e.g., scaling capacity by $\times 0.001$), so the test detects both natively active and conditionally bindable components.

\subsection{Structured Generation}
\label{sec:sg}

Direct code generation combines understanding, formalization, and implementation in a single output, so an error at any stage passes into the code unchecked. We decompose generation into four stages executed in a single LLM call:
\begin{equation}
\label{eq:cot}
x \;\xrightarrow{\;\text{understand}\;}\; \mathcal{U} \;\xrightarrow{\;\text{formalize}\;}\; \mathcal{M} \;\xrightarrow{\;\text{synthesize}\;}\; \hat{C} \;\xrightarrow{\;\text{verify}\;}\; C
\end{equation}

\paragraph{Stage~1 (Understand).}
Extract the objective direction, decision variables, constraints, and parameters from~$x$, producing a structured interpretation~$\mathcal{U}$ that makes implicit assumptions explicit.

\paragraph{Stage~2 (Formalize).}
Transform $\mathcal{U}$ into a mathematical specification $\mathcal{M} = (\mathcal{I}, \mathcal{P}, \mathcal{V}, \mathcal{C}, f)$: index sets~$\mathcal{I}$, parameters~$\mathcal{P}$, decision variables~$\mathcal{V}$ with explicit type reasoning, constraints~$\mathcal{C}$, and objective~$f$.
The prompt requires the LLM to decide from the physical context whether each variable is continuous, integer, or binary (e.g., ``can you order 2.7 pallets?''). This step targets a common failure in which models default to continuous variables for discrete decisions and thereby relax integrality.

\paragraph{Stage~3 (Synthesize).}
Generate executable Gurobi code $\hat{C}$ from $\mathcal{M}$ that reads all data through \texttt{data["key"]} dictionary lookups rather than hardcoded literals. L2 depends on this structure, because its perturbation tests modify parameter values in the runtime data dictionary. We therefore first prompt the LLM to extract all numerical parameters from $x$ into a JSON dictionary, and the generated code references this pre-loaded dictionary. If extraction fails (e.g., malformed JSON), we fall back to self-contained generation, in which the LLM embeds the data in the code; the perturbation engine detects and handles both patterns (Appendix~\ref{app:perturbation_strategy}).

\paragraph{Stage~4 (Verify Completeness).}
Cross-check the generated code $\hat{C}$ against the original problem $x$: are all cost and revenue terms in the objective, are all constraints implemented, and are all data values accessed correctly? If the check finds a discrepancy, the model fixes the code before returning the final $C$, so these errors are corrected before verification. Stage~4 is a \emph{grounded} completeness check rather than open-ended self-critique, because it compares the code with the original problem text $x$ through an explicit checklist. The negative results on LLM self-correction concern identifying reasoning errors \emph{without external feedback}~\citep{huang2024selfcorrection}. The external, non-LLM signal in ReLoop is L2's solver response (\S\ref{sec:l2}).

\subsection{Two-Layer Behavioral Verification}
\label{sec:verification}

Two principles govern the architecture.
(1)~\textbf{Only L1 blocks output.} L2 provides diagnostics but never discards a valid solution, so the pipeline always returns a solution that has passed L1.
(2)~\textbf{Conservative repair.} Only high-confidence issues (\textsc{Warning}) trigger repair; ambiguous signals (\textsc{Info}) are logged for reference only, so false positives do not trigger repairs that introduce regressions.
Table~\ref{tab:severity} summarizes the severity matrix.

\begin{table}[t]
\centering
\small
\caption{Severity matrix across verification layers. ($^\dagger$)~Reference only: marked ``do not fix.''}
\label{tab:severity}
\vspace{0.5em}
\begin{tabular}{@{}lllc@{}}
\toprule
\textbf{Layer} & \textbf{Check} & \textbf{Severity} & \textbf{Action} \\
\midrule
\multirow{4}{*}{\shortstack[l]{L1: Execution\\(blocking)}} 
              & Syntax / runtime error & \textsc{Fatal} & Regenerate \\
              & Infeasible & \textsc{Fatal} & Regen.\ (IIS) \\
              & Unbounded & \textsc{Fatal} & Regen.\ (ray) \\
              & Duality gap$^\dagger$ & \textsc{Info} & None \\
\midrule
\multirow{2}{*}{\shortstack[l]{L2: CPT\\(diagnostic)}} 
              & Missing ($r<5\%$) & \textsc{Warning} & Repair \\
              & Uncertain ($5\%{\leq}r{\leq}30\%$)$^\dagger$ & \textsc{Info} & None \\
\midrule
\multirow{2}{*}{\shortstack[l]{L2: OPT\\(diagnostic)}} 
              & Missing ($r<5\%$) & \textsc{Warning} & Repair \\
              & Uncertain ($5\%{\leq}r{\leq}30\%$)$^\dagger$ & \textsc{Info} & None \\
\bottomrule
\end{tabular}
\end{table}

\subsubsection{L1: Execution Verification (Blocking)}
\label{sec:l1}

L1 sequentially checks syntax (AST parsing), runtime completion, and solver status. It is the \emph{only blocking layer}: a \textsc{Fatal} result triggers regeneration with error feedback (up to $N$ attempts). When L1 passes, we record the baseline objective $z^* = \textsc{Opt}(C)$.

For \textsc{Infeasible} status, L1 computes the \emph{Irreducible Inconsistent Subsystem} (IIS), a minimal set of conflicting constraints, and passes their names and bounds to the LLM. The regeneration prompt therefore states which constraints conflict instead of only reporting infeasibility. For \textsc{Unbounded} status, L1 reports the variables of the unbounded ray to guide the correction of variable bounds.

\subsubsection{L2: Behavioral Testing (CPT + OPT)}
\label{sec:l2}

L2 detects missing formulation components, that is, constraints or objective terms that the problem description requires but the generated code lacks.
L2 avoids the self-consistency limit of LLM self-review~\citep{madaan2023selfrefine, huang2024selfcorrection} by assigning detection to the \emph{solver}. The LLM only extracts candidate constraints and objective terms, and detection uses parameter perturbation and the solver's response, which do not depend on the LLM's reasoning.

\paragraph{Detection mechanism.}
L2 implements Assumption~\ref{prop:perturb} with two symmetric modules.
\textbf{Constraint Presence Testing (CPT)} targets missing constraints: for each candidate extracted by the LLM (annotated with physical type: capacity, demand, etc.), CPT applies extreme perturbation to the governing parameter (capacity $\times 0.001$, demand $\times 100$, other $\times 0.01$).
\textbf{Objective Presence Testing (OPT)} targets missing cost/revenue terms with analogous perturbation (cost $\times 0.001$, revenue $\times 100$, other $\times 0.01$).

\paragraph{Graduated thresholds.}
Both modules measure the objective change ratio $r = |z' - z^*|\,/\,|z^*|$ and classify:
\begin{itemize}[leftmargin=1.8em,topsep=2pt,itemsep=1pt]
    \item $r < \tau_\ell = 5\%$: \textsc{Warning} (likely \emph{missing}), which triggers repair.
    \item $\tau_\ell \leq r \leq \tau_h = 30\%$: \textsc{Info} (\emph{uncertain}), which is logged without repair.
    \item $r > \tau_h$ or infeasibility: \textsc{Pass} (\emph{present}), which confirms that the component is active.
\end{itemize}
The asymmetric $[\tau_\ell, \tau_h]$ buffer favors under-detection over over-repair, because false positives trigger repairs that can cause regressions; accuracy varies by ${<}1\%$ across $\tau_\ell \in [1\%, 10\%]$ and $\tau_h \in [20\%, 50\%]$.

\paragraph{Detection scope.}
Table~\ref{tab:scope} lists which defects the mechanism can and cannot detect, and Appendix~\ref{app:defect_study} measures these boundaries on defects whose labels are known by construction.

\begin{table}[t]
\centering
\small
\caption{Detection scope of L2 behavioral testing.}
\label{tab:scope}
\vspace{0.5em}
\begin{tabular}{@{}ll@{}}
\toprule
\textbf{Defect} & \textbf{L2 behavior} \\
\midrule
Missing parameter-governed constraint & Detected (CPT) \\
Missing objective term & Detected (OPT) \\
Present-but-wrong constraint reusing the same parameters & Partial / not detected \\
Coefficient-magnitude errors & Not detected \\
Equivalent-looking but structurally different decompositions & Not detected \\
Errors inducing infeasibility or unboundedness & Handled by L1 (IIS / rays) \\
\bottomrule
\end{tabular}
\vspace{-0.5em}
\end{table}

\subsection{Diagnosis-Guided Repair}
\label{sec:dr}

All layers output \texttt{Diagnostic} objects in a common format that specify the severity, the target component, and the evidence (e.g., ``constraint \texttt{capacity\_limit}: perturbation $\times 0.001$ caused only 0.3\% objective change'').
\textsc{Fatal} triggers \emph{regeneration} with the error context (IIS constraints, unbounded rays); \textsc{Warning} triggers \emph{targeted repair} guided by the specific L2 diagnostics, and \textsc{Info} items are explicitly marked ``do not fix'' to prevent over-correction.

\paragraph{Safety and regression guards.}
Three mechanisms prevent regressions caused by repair. (1)~A \emph{safety check} blocks reassignment of data variables and dangerous imports, because we observed repair LLMs fabricating values that corrupt the problem. (2)~A \emph{regression guard} rolls back any repair that crashes, degrades the solver status, or shifts the objective by ${>}\tau_r = 4\%$. It mainly catches failures of the repair step (data fabrication, fundamental restructuring) rather than legitimate constraint additions, because adding a component that L2 flagged as a localized defect typically shifts the objective by less than $\tau_r$. (3)~A \emph{skip guard} bypasses repair when no \textsc{Warning} exists. Algorithm~\ref{alg:reloop} (Appendix~\ref{app:reloop_details}) states the procedure.

\section{RetailOpt-190 Benchmark}
\label{sec:benchmark}

\textbf{RetailOpt-190} is a benchmark for \emph{compositional constraint reasoning} (Table~\ref{tab:benchmark_comparison}).
All 190 instances are multi-period retail inventory optimization problems in which a retailer decides \emph{how much to order}, \emph{when to order}, and \emph{how to allocate} products across periods to minimize total cost subject to interacting operational constraints.
Each instance has 3 products, 5 locations, and 20 periods, except that one archetype uses 52 periods and one multi-echelon archetype uses 6 locations. An instance specifies 20+ parameters (demand forecasts, unit costs, storage capacities, shelf lives, lead times, supplier limits) and 10--30 constraints that must bind jointly.
The difficulty comes from constraint \emph{interactions} rather than linguistic complexity: for example, perishability limits effective inventory, which tightens storage constraints and forces earlier ordering that conflicts with lead times.
Each instance is self-contained, since the natural-language description specifies all parameters, and a correct solution requires translating them into a coherent mathematical program.
RetailOpt-190 has two roles in our study. First, it exposes the compositional feasibility--correctness gap, on which structured generation gives the largest gain. Second, its reference formulations support the injected-defect characterization of L2 in Appendix~\ref{app:defect_study}, which maps the boundaries of behavioral verification.

\begin{table}[t]
\centering
\small
\caption{Comparison with existing benchmarks.}
\label{tab:benchmark_comparison}
\vspace{0.5em}
\begin{tabular}{@{}lccccc@{}}
\toprule
\textbf{Benchmark} & \textbf{Domain} & \textbf{Inst.} & \textbf{Multi-period} & \textbf{Compositional} & \textbf{Data-Code Sep.} \\
\midrule
NL4Opt~\citep{ramamonjison2023nl4opt} & Generic & 289 & Few & \ding{55} & \ding{55} \\
MAMO~\citep{huang2024mamo} & Generic & $\sim$800 & Some & \ding{55} & \ding{55} \\
IndustryOR~\citep{huang2025orlm} & Mixed & 100 & Some & \ding{55} & \ding{55} \\
OptMATH~\citep{lu2025optmath} & Mixed & $\sim$360 & Some & \ding{55} & \ding{55} \\
\midrule
\textbf{RetailOpt-190} & \textbf{Retail} & 190 & \textbf{All} & \ding{51} & \ding{51} \\
\bottomrule
\end{tabular}
\end{table}

\paragraph{Construction.}
The benchmark spans eight scenario families (F1--F8) built by \emph{progressive composition} (Table~\ref{tab:families}). F1--F4 introduce core retail mechanisms (inventory dynamics, substitution, resource limits, demand variability) with increasing constraint interactions, and F5 tests feasibility boundaries with deliberately tight or infeasible configurations. F6 introduces discrete variables (MOQ, pack sizes) that require integrality reasoning, and F7--F8 extend the setting to multi-location networks that require multi-echelon coordination.
Each of the 38 archetypes has 5 numerical variants ($\pm$15\%, deterministic seeds), which gives 190 instances that vary the parameters while holding the structure fixed.
Ground-truth optimal values are computed by Gurobi~11.0~\citep{gurobi} applied to hand-crafted formulations for each archetype (Appendix~\ref{app:reference_milp}).

\begin{table}[t]
\centering
\small
\caption{RetailOpt-190 scenario families ($38 \times 5 = 190$ instances).}
\label{tab:families}
\vspace{0.5em}
\begin{tabular}{@{}clcl@{}}
\toprule
\textbf{ID} & \textbf{Family} & \textbf{Arch.} & \textbf{Key Mechanisms} \\
\midrule
F1 & Core Operations & 4 & Multi-period inventory, perishability, lost sales \\
F2 & Assortment \& Substitution & 6 & Product substitution, promotions, price bands \\
F3 & Resource Constraints & 4 & Storage bottleneck, supply limits, volumetric \\
F4 & Demand Dynamics & 6 & Demand surge, supply risk, quality holds \\
F5 & Feasibility Stress & 4 & Impossible demand, storage overflow, service traps \\
F6 & Discrete Logistics & 4 & Lead time, MOQ, pack size, fixed order cost \\
F7 & Network \& Multi-Echelon & 6 & Transshipment, hub-spoke, multi-sourcing \\
F8 & Omni-channel & 4 & Reverse logistics, labor, ship-from-store \\
\bottomrule
\end{tabular}
\end{table}

\paragraph{Evaluation.}
Each instance is presented as a \emph{data-embedded} prompt with the JSON data inline, as in prior benchmarks~\citep{ramamonjison2023nl4opt, huang2024mamo, huang2025orlm}.
The model generates self-contained Python code, which is executed with a 60-second solver time limit.
An instance is correct if its feasibility status matches the ground truth and $|y_{\text{pred}} - y_{\text{ref}}| / |y_{\text{ref}}| < \epsilon$.
We report accuracy at two tiers. The \emph{strict} tier ($\epsilon = 10^{-4}$) requires near-identical objectives, so all constraints and coefficients must be correct. The \emph{practical} tier ($\epsilon = 10^{-2}$) also accepts near-correct solutions with minor coefficient discrepancies, that is, formulations that are structurally right but numerically imprecise.
The MIP family~F6 uses $\epsilon = 10^{-2}$ for both tiers because of the solver tolerance.
Cross-benchmark experiments use $\epsilon = 10^{-6}$, following~\citet{chen2025solverinformedrl}.

\section{Experiments and Results}
\label{sec:experiments}

\subsection{Setup}
\label{sec:exp_setup}

\paragraph{Models.}
We evaluate five models from three paradigms.
The \emph{foundation LLMs} are Claude~Opus~4.6 (Anthropic's frontier model), DeepSeek-V3.2~\citep{deepseek2024v3} (671B MoE), and Qwen3-32B~\citep{yang2025qwen3} (dense 32B).
The \emph{offline SFT} model is OptMATH-Qwen2.5-32B~\citep{lu2025optmath}, fine-tuned on $\sim$17K curated optimization problems, and the \emph{online RL} model is SIRL-Qwen2.5-32B~\citep{chen2025solverinformedrl}, trained with solver execution status as a verifiable reward.
Each model is evaluated under three configurations:
\emph{Base} (direct generation, or the model's own format),
\emph{CoT} (structured chain-of-thought), and
\emph{ReLoop} (CoT + L1--L2 verification with repair, $N\!=\!3$).
All runs use greedy decoding (temperature~0, pass@1).
ReLoop adds ${\sim}3\times$ the base cost in LLM tokens, and solver calls complete in seconds (Appendix~\ref{app:reloop_details}).

\paragraph{Benchmarks.}
RetailOpt-190 (\S\ref{sec:benchmark}),
MAMO-ComplexLP (203 instances from the challenging subset of MAMO)~\citep{huang2024mamo},
and IndustryOR (100 real-world instances)~\citep{huang2025orlm}.

\paragraph{Metrics.}
On RetailOpt-190: \textbf{Exec\%} (fraction producing a solver-feasible solution), \textbf{Acc\%($\epsilon\!=\!10^{-4}$)} (strict formulation correctness), and \textbf{Acc\%($\epsilon\!=\!10^{-2}$)} (practical accuracy).
Cross-benchmark: \textbf{Acc\%($\epsilon\!=\!10^{-6}$)} following~\citet{chen2025solverinformedrl}.

\subsection{Main Results on RetailOpt-190}
\label{sec:main_results}

\begin{table*}[t]
\centering
\small
\caption{Main results on RetailOpt-190 (pass@1, greedy decoding, $N\!=\!3$).}
\label{tab:main}
\vspace{0.5em}
\setlength{\tabcolsep}{3.2pt}
\begin{tabular}{@{}ll ccc ccc ccc @{}}
\toprule
& & \multicolumn{3}{c}{\textbf{Exec\%}} & \multicolumn{3}{c}{\textbf{Acc\%($\epsilon\!=\!10^{-4}$)}} & \multicolumn{3}{c}{\textbf{Acc\%($\epsilon\!=\!10^{-2}$)}} \\
\cmidrule(lr){3-5} \cmidrule(lr){6-8} \cmidrule(lr){9-11}
\textbf{Type} & \textbf{Model} & Base & CoT & ReLoop & Base & CoT & ReLoop & Base & CoT & ReLoop \\
\midrule
\multirow{3}{*}{\textit{Foundation}} 
  & Claude~Opus~4.6  & 72.1 & 93.7 & \textbf{100.0} & 22.6 & 31.1 & \textbf{31.1} & 26.8 & 34.7 & \textbf{35.3} \\
  & DeepSeek-V3.2    & 91.1 & 53.2 & \textbf{97.4} & 0.5 & 3.7 & \textbf{5.8} & 3.7 & 5.8 & \textbf{11.1} \\
  & Qwen3-32B        & 0.0 & 0.0 & \textbf{2.1} & 0.0 & 0.0 & 0.0 & 0.0 & 0.0 & 0.0 \\
\midrule
\textit{Offline SFT}
  & OptMATH-32B      & 2.6 & 2.6 & \textbf{17.9} & 0.0 & 0.0 & 0.0 & 0.0 & 0.0 & \textbf{0.5} \\
\midrule
\textit{Online RL}
  & SIRL-32B         & 0.0 & 0.0 & \textbf{1.6} & 0.0 & 0.0 & 0.0 & 0.0 & 0.0 & 0.0 \\
\bottomrule
\end{tabular}
\end{table*}

Table~\ref{tab:main} shows that \textbf{feasibility is a poor proxy for correctness}. DeepSeek reaches 91.1\% solver feasibility but only 0.5\% correctness, a 90-point gap, and a large gap persists after ReLoop (Claude: 100\% Exec, 31.1\% Acc, so two-thirds of the runs remain silent failures).

The models fail in different ways, and ReLoop preserves or improves every foundation-model metric relative to Base.
On Claude, CoT gives the largest accuracy gain (+8.5pp). On DeepSeek, CoT lowers execution (91.1\%$\to$53.2\%) because the four-stage decomposition produces intermediate mathematical notation that does not translate into valid Gurobi syntax. L1 fully recovers this loss through its \textsc{Fatal} feedback (syntax tracebacks for parser errors, IIS for infeasibility, unbounded rays for unboundedness).
Claude's accuracy stays at 31.1\% because most of the remaining failures are \emph{structural} silent failures: problem decompositions that differ from the reference yet are internally consistent, which perturbation cannot detect.
The 32B models reach near-zero accuracy, consistent with RetailOpt-190's compositional structure (20+ interacting constraints across multi-period, multi-product, multi-location dimensions) exceeding their reasoning capacity. ReLoop still improves their execution (OptMATH: 2.6\%$\to$17.9\%; SIRL: 0.0\%$\to$1.6\%), which suggests that strong models benefit from CoT structuring and weaker models from L1 crash recovery.
L2 adds gains for the foundation models without lowering any foundation-model metric (its largest effect is on MAMO's localized defects, \S\ref{sec:ablation}); the non-blocking, regression-guarded design limits the losses from incorrect repairs.
For Claude, a paired exact sign test supports the CoT effect (20 corrected vs.\ 4 regressed, $p \approx 0.002$). Because the 190 instances are 38 archetypes $\times$ 5 numerical variants, we also run an archetype-level cluster bootstrap (38 clusters, 10{,}000 resamples), which gives a 95\% confidence interval of $[1.6, 16.8]$pp for the CoT gain.

The gap between the strict and practical tiers ($\epsilon\!=\!10^{-4}$ vs.\ $10^{-2}$) is also informative: DeepSeek's larger spread (5.8\% vs.\ 11.1\%) suggests near-correct solutions with coefficient errors, whereas Claude's smaller spread (31.1\% vs.\ 35.3\%) suggests mostly structural errors.

\subsection{Cross-Benchmark Generalization}
\label{sec:cross_benchmark}

\begin{table}[t]
\centering
\small
\caption{Cross-benchmark generalization (Acc\%, $\epsilon\!=\!10^{-6}$, pass@1).}
\label{tab:cross}
\vspace{0.5em}
\setlength{\tabcolsep}{3pt}
\begin{tabular}{@{}ll ccc ccc @{}}
\toprule
& & \multicolumn{3}{c}{\textbf{MAMO-ComplexLP}} & \multicolumn{3}{c}{\textbf{IndustryOR}} \\
\cmidrule(lr){3-5} \cmidrule(lr){6-8}
\textbf{Type} & \textbf{Model} & Base & CoT & +ReLoop & Base & CoT & +ReLoop \\
\midrule
\multirow{3}{*}{\textit{Foundation}} 
  & Claude~Opus~4.6  & 70.4 & 73.9 & \textbf{79.8} & 66.0 & 66.0 & \textbf{68.0} \\
  & DeepSeek-V3.2    & 60.1 & 59.6 & \textbf{62.6} & 50.0 & 54.0 & \textbf{62.0} \\
  & Qwen3-32B        & 40.4 & 37.4 & \textbf{46.3} & 43.0 & 43.0 & \textbf{46.0} \\
\midrule
\textit{Offline SFT}
  & OptMATH-32B      & \textbf{56.2} & 30.0 & 31.0 & \textbf{34.0} & 31.0 & \textbf{34.0} \\
\midrule
\textit{Online RL}
  & SIRL-32B         & 53.2 & 46.8 & \textbf{54.2} & 40.0 & 40.0 & \textbf{43.0} \\
\bottomrule
\end{tabular}
\end{table}

To test generalization beyond the retail domain of RetailOpt-190, we evaluate on two external benchmarks, neither of which was used during development (Table~\ref{tab:cross}; composition audited in Appendix~\ref{app:composition}). MAMO-ComplexLP contains LP/MILP problems with well-isolated constraints ($\sim$459 tokens on average), and IndustryOR contains real-world problems that require multi-step reasoning.

ReLoop improves all three foundation models and SIRL on both benchmarks without domain-specific tuning.
SIRL, trained with solver execution feedback via reinforcement learning, also improves: CoT alone lowers its MAMO accuracy, but unlike OptMATH it ends above Base once verification and repair are added.
The exception is OptMATH on MAMO, whose accuracy falls from 56.2\% (Base) to 30.0\% under CoT: the SFT model's rigid ``problem$\to$code'' generation pattern does not accommodate our four-stage reasoning template, so 84 instances crash and 65 previously correct solutions become incorrect.\footnote{The high Base score on MAMO may also reflect training--evaluation overlap, since OptMATH's ${\sim}$17K curated corpus could include publicly available benchmark problems. On IndustryOR, the Base--CoT gap is much smaller (34.0\%$\to$31.0\%).}
All three 32B models stay well below frontier performance (best: 46.3\% on MAMO, 46.0\% on IndustryOR), which suggests that 32B-scale models lack the reasoning capacity for industrial optimization problems of this complexity, even with verification.
The size of the gain depends on the error structure. DeepSeek gains +12.0pp on IndustryOR, +4.0pp from CoT and +8.0pp from L1, with no marginal gain from L2 (50.0/54.0/62.0/62.0 across Direct/+CoT/+CoT+L1/+ReLoop; Claude: 66.0/66.0/68.0/68.0). Claude gains +4.4pp on MAMO through L2 behavioral testing (11 corrected vs.\ 2 regressed, $p \approx 0.02$), and all 11 corrections restore integrality that continuous relaxation had dropped (Appendix~\ref{app:audit}).

\subsection{Ablation Studies}
\label{sec:ablation}

\begin{table*}[t]
\centering
\small
\caption{Ablation across benchmarks (pass@1). Each row adds one component; L2 = CPT + OPT behavioral testing.}
\label{tab:ablation}
\vspace{0.5em}
\setlength{\tabcolsep}{2.8pt}
\begin{tabular}{@{}l ccc ccc cc cc @{}}
\toprule
& \multicolumn{6}{c}{\textbf{RetailOpt-190}} & \multicolumn{4}{c}{\textbf{MAMO-ComplexLP} ($\epsilon\!=\!10^{-6}$)} \\
\cmidrule(lr){2-7} \cmidrule(lr){8-11}
& \multicolumn{3}{c}{Claude~Opus~4.6} & \multicolumn{3}{c}{DeepSeek-V3.2} & \multicolumn{2}{c}{Claude~Opus~4.6} & \multicolumn{2}{c}{DeepSeek-V3.2} \\
\cmidrule(lr){2-4} \cmidrule(lr){5-7} \cmidrule(lr){8-9} \cmidrule(lr){10-11}
\textbf{Config} & \textbf{Exec} & \textbf{Acc$_{10^{-4}}$} & \textbf{Acc$_{10^{-2}}$} & \textbf{Exec} & \textbf{Acc$_{10^{-4}}$} & \textbf{Acc$_{10^{-2}}$} & \textbf{Exec} & \textbf{Acc} & \textbf{Exec} & \textbf{Acc} \\
\midrule
Direct          & 72.1 & 22.6 & 26.8 & 91.1 & 0.5 & 3.7 & 94.1 & 70.4 & 93.6 & 60.1 \\
+CoT            & 93.7 & 31.1 & 34.7 & 53.2 & 3.7 & 5.8 & 95.6 & 73.9 & 87.7 & 59.6 \\
+CoT+L1         & 99.5 & 31.1 & 35.3 & 97.4 & 5.8 & 10.5 & 98.0 & 75.4 & 88.7 & 60.6 \\
+CoT+L1+L2      & \textbf{100.0} & \textbf{31.1} & \textbf{35.3} & \textbf{97.4} & \textbf{5.8} & \textbf{11.1} & \textbf{98.0} & \textbf{79.8} & \textbf{88.7} & \textbf{62.6} \\
\bottomrule
\end{tabular}
\end{table*}

Table~\ref{tab:ablation} isolates each component's marginal contribution across two benchmarks.

\textbf{Direct $\to$ +CoT.}
On Claude, structured generation gives the largest accuracy gain (+8.5pp; 20 corrected, 4 regressed). We attribute this gain to explicit structured reasoning in general; separating the OR-specific staging from generic step-by-step reasoning would require a vanilla chain-of-thought control, which we leave to future work.
On DeepSeek, CoT lowers execution (91.1\%$\to$53.2\%) because the structured format produces invalid code, which motivates L1 as a dedicated recovery layer.

\textbf{+CoT $\to$ +CoT+L1.}
L1 accounts for most of the execution recovery (+44.2pp Exec for DeepSeek, +5.8pp for Claude) and also improves accuracy.
When L1 detects infeasibility, the IIS identifies a minimal set of conflicting constraints (e.g., ``\texttt{capacity\_limit} conflicts with \texttt{demand\_fulfillment}''), so the LLM receives the specific conflict and can reformulate the affected constraints instead of retrying without information.

\textbf{+CoT+L1 $\to$ +CoT+L1+L2.}
L2's contribution depends on the error structure.
On MAMO, L2 is the \emph{largest single accuracy contributor} for Claude (+4.4pp; 11 corrected, 2 regressed) and adds +2.0pp for DeepSeek (4 corrected, 0 regressed), because the simpler problem structures produce \emph{localized errors}, that is, missing constraints or objective terms, which are the defects that perturbation testing detects.
On RetailOpt-190, L2 adds execution recovery (Claude: 99.5\%$\to$100.0\%) and practical accuracy (DeepSeek: 10.5\%$\to$11.1\% at $\epsilon\!=\!10^{-2}$), but strict accuracy ($\epsilon\!=\!10^{-4}$) is unchanged because the errors are mostly structural: incorrect decompositions that still respond plausibly to perturbation. Two appendix analyses support this reading. In an expert-verified audit of 55 residual failures from both benchmarks, 42 of the 43 true modeling errors are structural (Appendix~\ref{app:audit}). In an injected-defect study, the detector has 99.2\% recall on 240 construction-labeled knockouts and a 26.7\% false-positive rate on candidates in intact code (Appendix~\ref{app:defect_study}).
On IndustryOR, the relative objective deviation $|z_{\text{pred}} - z_{\text{ref}}|/|z_{\text{ref}}|$ over all non-crashed instances is bimodal: 34\% deviate by less than 1\% (small coefficient differences that perturbation cannot detect), and 47\% deviate by more than 10\% (structural misunderstandings that 3 repair iterations do not fix), so almost no instances fall in the range that repair can correct.
Per-family results are in Appendix~\ref{app:family_analysis}.

\section{Conclusion}
\label{sec:conclusion}

ReLoop addresses silent failures in LLM-generated optimization code with \emph{structured generation}, which reduces formulation errors during generation, and \emph{behavioral verification}, which detects the remaining errors through solver-based perturbation without ground truth.
The cross-benchmark ablation shows that the two mechanisms address different error structures: structured generation gives the largest gain on the compositional problems of RetailOpt-190 (\textbf{+8.5pp} accuracy on Claude Opus~4.6), and behavioral verification gives its largest gain on the localized defects of MAMO-ComplexLP (\textbf{+4.4pp} on Claude).
Relative to Base, ReLoop preserves or improves every foundation-model metric. The 90-point feasibility--correctness gap shows that solver feasibility does not certify an LLM-generated model, so semantic verification is needed before such models are deployed.

\paragraph{Limitations.}
\emph{Supported formulation class.} The current implementation targets (MI)LPs whose parameters have nonnegative physical meaning (capacities, demands, costs, resource coefficients). Signed or near-zero coefficients rely on the absolute-change fallback of Assumption~\ref{prop:perturb}, and nonlinear or piecewise objectives and stochastic or chance-constrained models are out of scope. Robustness to the multiplier is established only for shrink-direction perturbations (Appendix~\ref{app:defect_study}), because RetailOpt-190 is a cost-minimization benchmark with no test population for the inflate direction.
\emph{Format compatibility.} CoT disrupts the learned output patterns of SFT models (84 crashes and 65 regressions for OptMATH on MAMO).
\emph{Shared extraction.} L2 uses the generating LLM to extract candidates, so extraction and generation errors may be correlated. The oracle-extraction study separates extraction error from detection error (Appendix~\ref{app:defect_study}), and the symmetric regression threshold $\tau_r$ is calibrated for localized defects, so repairs of major structural omissions are rolled back and those instances keep the L1 result.
\emph{Detection boundaries.} Coefficient-magnitude errors, structurally different decompositions that look equivalent, and present-but-wrong constraints that share parameters are beyond the scope of L2 (Table~\ref{tab:scope}). Solution-level change could be used as a secondary \textsc{Info} signal, and an IIS-based test that separates intrinsic from formulation-induced infeasibility would extend ReLoop to problems whose correct answer may be ``infeasible.''

\begin{ack}
This research is supported by the National Research Foundation, Singapore and Infocomm Media Development Authority under its Trust Tech Funding Initiative and the National Natural Science Foundation of China. Any opinions, findings and conclusions or recommendations expressed in this material are those of the author(s) and do not reflect the views of National Research Foundation, Singapore and Infocomm Media Development Authority. The DTC grant number is DTC-IGC-07 (project: Multi-Party Evaluation: Blockchain-Enabled AI Safety and Verification Framework), and the National Natural Science Foundation of China grant numbers are U25A20450 and 62571374. The authors declare no competing interests.
\end{ack}

\small

\bibliographystyle{unsrtnat}  
\bibliography{references}
\normalsize

\newpage

\appendix

\section{Reference MILP Formulation}
\label{app:reference_milp}

This appendix gives the modular MILP that defines the semantic ground truth for RetailOpt-190.
We present the notation and decision variables, the objective function, and all constraint families with explicit equations,
followed by the \emph{design rationale} for modeling choices that prevent common LLM errors.

\subsection{Notation}

\paragraph{Sets and Indices.}
\begin{itemize}[leftmargin=*, nosep]
    \item $\mathcal{P}$: set of products (SKUs); $\mathcal{L}$: set of locations (distribution centers/stores)
    \item $\mathcal{T}=\{1,\dots,T\}$: discrete planning periods (weeks)
    \item $\mathcal{K}_p=\{1,\dots,\text{SL}_p\}$: remaining shelf-life buckets for product $p$
    \item $\mathcal{E}^{\text{sub}}\subseteq\mathcal{P}\times\mathcal{P}$: directed substitution arcs (upward substitution)
    \item $\mathcal{E}^{\text{tr}}\subseteq\mathcal{L}\times\mathcal{L}$: directed transshipment arcs
    \item $\mathcal{N}^{\text{out}}_p = \{p' : (p,p')\in\mathcal{E}^{\text{sub}}\}$: products that can serve $p$'s demand
    \item $\mathcal{N}^{\text{in}}_p = \{p' : (p',p)\in\mathcal{E}^{\text{sub}}\}$: products whose demand $p$ can serve
\end{itemize}

\paragraph{Parameters.}
\begin{itemize}[leftmargin=*, nosep]
    \item $d_{p,l,t}$: demand for product $p$ at location $l$ in period $t$, computed as $d_{p,l,t} = \texttt{demand\_curve}[p][t{-}1] \times \texttt{demand\_share}[l]$
    \item $\text{SL}_p$: shelf life of product $p$ in periods
    \item $\text{LT}_p$: order lead time for product $p$ (periods between placement and receipt)
    \item $\bar{Q}_{p,t}$: production/procurement capacity for product $p$ in period $t$
    \item $\bar{C}_l$: cold storage capacity at location $l$ (volume units)
    \item $\gamma_p$: cold storage usage coefficient per unit of product $p$
    \item $\bar{H}_{l,t}$: labor capacity at location $l$ in period $t$ (hours)
    \item $h_p$: labor usage per unit of product $p$ handled
    \item $\rho_p$: return rate, the fraction of period-$t$ sales returned in period $t{+}1$
    \item $c^{\text{buy}}_p, c^{\text{inv}}_p, c^{\text{waste}}_p, c^{\text{ls}}_p$: purchasing, holding, waste, and lost-sales costs
    \item $c^{\text{fix}}$: fixed ordering cost per order placed; $c^{\text{tr}}$: transshipment cost per unit
    \item $\text{MOQ}$: minimum order quantity; $\text{PS}$: pack size
    \item $B$: per-period purchasing budget (if active); $\omega$: maximum waste fraction (if active)
\end{itemize}

\paragraph{Key Convention.}
Remaining-life index $k=\text{SL}_p$ denotes the \emph{freshest} inventory; $k=1$ denotes inventory expiring at the end of the current period.
This FIFO-compatible convention matters because LLMs frequently reverse the ordering, which produces aging dynamics in which fresh inventory expires immediately (\S\ref{app:design_rationale}, D2).

\subsection{Decision Variables}

Table~\ref{tab:variables} lists all decision variables in the reference MILP.
The formulation uses a \emph{modular activation} pattern: the five core variables ($I$, $y$, $Q$, $W$, $L$) are always created, while optional variables ($S$, $X$, $z$, $n$) are instantiated only when the corresponding mechanism appears in the instance JSON.
This follows the solver implementation, where variable creation depends on field checks such as \texttt{pack\_size > 1} or \texttt{len(sub\_edges) > 0}.

\begin{table}[ht]
\centering
\small
\caption{Decision variables. Variables marked with $^\dagger$ are created only when the corresponding mechanism is active (i.e., the relevant JSON field is non-default).}
\label{tab:variables}
\begin{tabular}{@{}lll@{}}
\toprule
\textbf{Variable} & \textbf{Domain} & \textbf{Interpretation} \\
\midrule
$I_{p,l,t,k}$ & $\mathbb{R}_{\ge 0}$ & Start-of-period inventory of product $p$ at location $l$ \\
               &                       & in period $t$ with $k$ periods of remaining life \\
$y_{p,l,t,k}$ & $\mathbb{R}_{\ge 0}$ & Sales from remaining-life bucket $k$ in period $t$ \\
$Q_{p,l,t}$   & $\mathbb{R}_{\ge 0}$ & Order quantity placed for product $p$ at location $l$ in period $t$ \\
$W_{p,l,t}$   & $\mathbb{R}_{\ge 0}$ & Waste (expired units) at end of period $t$ \\
$L_{p,l,t}$   & $\mathbb{R}_{\ge 0}$ & Lost sales (unmet demand) in period $t$ \\
\midrule
$S_{p \to p',l,t}^\dagger$ & $\mathbb{R}_{\ge 0}$ & Substitution flow: units of $p$'s demand served by $p'$'s inventory \\
                            &                       & (created iff $\mathcal{E}^{\text{sub}} \neq \emptyset$) \\
$X_{p,l \to l',t}^\dagger$ & $\mathbb{R}_{\ge 0}$ & Transshipment flow from $l$ to $l'$ \\
                            &                       & (created iff $\mathcal{E}^{\text{tr}} \neq \emptyset$) \\
$z_{p,l,t}^\dagger$        & $\{0,1\}$            & Binary order trigger \\
                            &                       & (created iff $\text{MOQ} > 0$ or $c^{\text{fix}} > 0$) \\
$n_{p,l,t}^\dagger$        & $\mathbb{Z}_{\ge 0}$ & Integer pack count (created iff $\text{PS} > 1$) \\
\bottomrule
\end{tabular}
\end{table}

\subsection{Objective Function}

The objective minimizes total supply chain cost over the planning horizon:
\begin{equation}
\label{eq:objective}
\min \sum_{p \in \mathcal{P}} \sum_{l \in \mathcal{L}} \sum_{t \in \mathcal{T}} \bigg[
    \underbrace{c^{\text{buy}}_p \, Q_{p,l,t}}_{\text{purchasing}}
    + \underbrace{c^{\text{inv}}_p \sum_{k=2}^{\text{SL}_p} \big(I_{p,l,t,k} - y_{p,l,t,k}\big)}_{\text{holding}}
    + \underbrace{c^{\text{waste}}_p \, W_{p,l,t}}_{\text{waste}}
    + \underbrace{c^{\text{ls}}_p \, L_{p,l,t}}_{\text{lost sales}}
    \bigg] + \Phi^{\text{opt}}
\end{equation}
where $\Phi^{\text{opt}}$ collects optional cost terms activated by specific mechanism flags:
\begin{equation}
\label{eq:optional_costs}
\Phi^{\text{opt}} = \sum_{p,l,t} c^{\text{fix}} \, z_{p,l,t} \cdot \mathbb{1}[c^{\text{fix}}>0]
\;+\; \sum_{p} \sum_{(l,l') \in \mathcal{E}^{\text{tr}}} \sum_{t} c^{\text{tr}} \, X_{p,l \to l',t}
\end{equation}

\paragraph{Design Note (Holding Cost Scope).}
Holding cost applies only to buckets $k \ge 2$ because bucket $k{=}1$ contains inventory that either sells or expires within the current period, so it \emph{cannot} be carried into the next period.
LLMs frequently apply holding cost to all buckets $k \ge 1$, which double-counts the waste penalty on expiring units (D5 in \S\ref{app:design_rationale}).
The end-of-period inventory in each bucket is $I_{p,l,t,k} - y_{p,l,t,k}$, representing units that survived sales and will age into the next period.

\subsection{Core Constraints}
\label{app:core_constraints}

The reference formulation enforces six core constraint families that govern
inventory flow across products, locations, and time periods
(Table~\ref{tab:constraints}). Together they form a \emph{closed conservation
system}: every unit entering the system (via ordering) must exit through
exactly one of four channels: sales, aging into the next period,
expiration, or lost sales. This accounting is what makes
silent failures detectable: if an LLM omits or reverses any single
constraint, the conservation structure breaks, and the resulting solutions
are solver-feasible but semantically inconsistent.

\begin{table}[ht]
\centering
\small
\caption{Core constraint families with explicit equations and design rationale. Each rationale column explains the specific modeling choice; see \S\ref{app:design_rationale} for detailed justification.}
\label{tab:constraints}
\setlength{\tabcolsep}{3pt}
\begin{tabular}{@{}p{1.8cm}p{6.5cm}p{4.5cm}@{}}
\toprule
\textbf{Family} & \textbf{Constraint} & \textbf{Design Rationale} \\
\midrule
C1.\ Initialization &
$I_{p,l,1,k} = 0 \quad \forall\, k < \text{SL}_p$ \newline
$I_{p,l,1,\text{SL}_p} = A_{p,l,1}$
&
Without zero initialization for non-fresh buckets, unbounded phantom inventory yields objective $\approx 0$ (D1). \\
\addlinespace[4pt]

C2.\ Fresh inflow &
$I_{p,l,t,\text{SL}_p} = A_{p,l,t} + \Delta^{\text{tr}}_{p,l,t} + R_{p,l,t}$ \newline
where $A_{p,l,t} = Q_{p,l,t-\text{LT}_p}$ if $t{-}\text{LT}_p \ge 1$, else $0$;\newline
$\Delta^{\text{tr}}_{p,l,t} = \sum_{l': (l',l) \in \mathcal{E}^{\text{tr}}} X_{p,l' \to l,t} - \sum_{l': (l,l') \in \mathcal{E}^{\text{tr}}} X_{p,l \to l',t}$;\newline
$R_{p,l,t} = \rho_p \sum_{k} y_{p,l,t-1,k}$ if $t > 1$, else $0$.
&
Fresh inventory enters \emph{only} via ordering, transshipment, and returns. LLMs that subtract sales in this equation effectively double-count demand fulfillment. \\
\addlinespace[4pt]

C3.\ Aging dynamics &
$I_{p,l,t+1,k} = I_{p,l,t,k+1} - y_{p,l,t,k+1}$ \newline
$\quad \forall\, t \in \{1,\dots,T{-}1\},\; k \in \{1,\dots,\text{SL}_p{-}1\}$
&
Remaining life \emph{decrements} each period: bucket $k{+}1$ today becomes bucket $k$ tomorrow, minus sales. Incrementing $k$ instead causes fresh stock to expire immediately (D2). \\
\addlinespace[4pt]

C4.\ Expiration &
$W_{p,l,t} = I_{p,l,t,1} - y_{p,l,t,1}$
&
Only bucket $k{=}1$ can expire. Unsold units from the oldest bucket become waste; waste is \emph{not} carried forward. \\
\addlinespace[4pt]

C5.\ Sales availability &
$y_{p,l,t,k} \le I_{p,l,t,k} \quad \forall\, k$
&
Cannot sell more than on-hand inventory in each remaining-life bucket. Without this, the solver can create phantom sales. \\
\addlinespace[4pt]

C6.\ Demand conservation &
$\sum_{k=1}^{\text{SL}_p} y_{p,l,t,k} + L_{p,l,t} = d_{p,l,t} + S^{\text{in}}_{p,l,t} - S^{\text{out}}_{p,l,t}$ \newline
with $S^{\text{out}}_{p,l,t} = \sum_{p' \in \mathcal{N}^{\text{out}}_p} S_{p \to p',l,t}$, \newline
$S^{\text{in}}_{p,l,t} = \sum_{p' \in \mathcal{N}^{\text{in}}_p} S_{p' \to p,l,t}$, \newline
$S^{\text{out}}_{p,l,t} \le d_{p,l,t}$ (demand-route bound), \newline
and $S^{\text{in}}_{p,l,t} \le \sum_{k=1}^{\text{SL}_p} y_{p,l,t,k}$ (inventory-backed substitution).
&
Reversing edge direction is a common modeling mistake (D4): $S_{p \to p'}$ means $p$'s demand \emph{exported to} $p'$'s inventory, not the reverse. The demand-route and inventory-backed bounds together prevent unbounded substitution flow. \\
\bottomrule
\end{tabular}
\end{table}

\subsection{Capacity and Resource Constraints}
\label{app:capacity_constraints}

In addition to the six core inventory-flow constraints, the formulation enforces a set of capacity and resource limits. These are modular: each constraint is active only when the corresponding JSON parameter takes a non-trivial value.

\paragraph{Production Capacity.}
Each product's total inflow across all locations is bounded by a time-varying capacity:
\begin{equation}
\sum_{l \in \mathcal{L}} A_{p,l,t} \;\le\; \bar{Q}_{p,t}, \qquad \forall\, p \in \mathcal{P},\; t \in \mathcal{T}
\end{equation}
where $A_{p,l,t}$ denotes the arrival quantity (accounting for lead time). This constraint is applied to \emph{delivered} inflow, ensuring lead-time consistency: an order placed in period $t$ arrives in period $t + \text{LT}_p$ and counts against the capacity of the \emph{arrival} period.

\paragraph{Storage Capacity.}
Total volume-weighted on-hand inventory at each location must not exceed cold storage capacity:
\begin{equation}
\sum_{p \in \mathcal{P}} \gamma_p \sum_{k=1}^{\text{SL}_p} I_{p,l,t,k} \;\le\; \bar{C}_l, \qquad \forall\, l \in \mathcal{L},\; t \in \mathcal{T}
\end{equation}
Storage usage is measured on \emph{start-of-period} inventory (before sales), because this is the physical space occupied when replenishment decisions are made. Product-specific usage coefficients $\gamma_p$ model volumetric heterogeneity (e.g., bulky premium items in F3 archetypes).

\paragraph{Labor Capacity.}
Total labor consumed for sales handling at each location is bounded:
\begin{equation}
\sum_{p \in \mathcal{P}} h_p \sum_{k=1}^{\text{SL}_p} y_{p,l,t,k} \;\le\; \bar{H}_{l,t}, \qquad \forall\, l \in \mathcal{L},\; t \in \mathcal{T}
\end{equation}
Labor is modeled as proportional to units sold (picked and packed). In many archetypes, $\bar{H}_{l,t}$ is set to a large value so the constraint does not bind; it becomes active in F8 archetypes where omni-channel operations increase per-unit labor intensity.

\paragraph{Per-Period Budget.}
When $B \neq \texttt{null}$, total purchasing spend in each period is capped:
\begin{equation}
\sum_{p \in \mathcal{P}} \sum_{l \in \mathcal{L}} c^{\text{buy}}_p \, Q_{p,l,t} + c^{\text{fix}} \sum_{p,l} z_{p,l,t} \;\le\; B, \qquad \forall\, t \in \mathcal{T}
\end{equation}

\paragraph{Global Waste Cap.}
When $\omega \neq \texttt{null}$, total waste over the horizon is limited as a fraction of total demand:
\begin{equation}
\sum_{p,l,t} W_{p,l,t} \;\le\; \omega \cdot \sum_{p,l,t} d_{p,l,t}
\end{equation}

\subsection{Discrete Procurement Constraints}
\label{app:discrete_constraints}

Family F6 activates integer procurement logic. These constraints introduce mixed-integer structure via binary and integer variables.

\paragraph{Minimum Order Quantity (MOQ).}
When $\text{MOQ} > 0$, each order must be either zero or at least $\text{MOQ}$ units:
\begin{align}
Q_{p,l,t} &\le M \cdot z_{p,l,t} \label{eq:moq_ub} \\
Q_{p,l,t} &\ge \text{MOQ} \cdot z_{p,l,t} \label{eq:moq_lb}
\end{align}
where $M = 10^6$ is a big-M constant and $z_{p,l,t} \in \{0,1\}$ is the binary order trigger. The big-M formulation is standard; conditional constraints are not directly expressible in MILP and require this linearization.

\paragraph{Pack Size.}
When $\text{PS} > 1$, order quantities must be integer multiples of the pack size:
\begin{equation}
Q_{p,l,t} = \text{PS} \cdot n_{p,l,t}, \qquad n_{p,l,t} \in \mathbb{Z}_{\ge 0}
\end{equation}

\paragraph{Fixed Ordering Cost.}
When $c^{\text{fix}} > 0$, the binary trigger $z_{p,l,t}$ is activated (via Eq.~\ref{eq:moq_ub}) and the fixed cost $c^{\text{fix}} \cdot z_{p,l,t}$ enters the objective.

\subsection{Formulation Design Rationale}
\label{app:design_rationale}

Several formulation choices in RetailOpt-190 need an explicit justification, because in each case a na\"ive implementation produces a solver-feasible but semantically incorrect solution.

\begin{enumerate}[leftmargin=*, itemsep=6pt, label=\textbf{D\arabic*.}]
    \item \textbf{Explicit zero initialization ($I_{p,l,1,k}=0$ for $k < \text{SL}_p$).}
    The reference formulation initializes all non-fresh inventory buckets to zero in period~1. Without this, the model admits unbounded phantom inventory in non-fresh buckets, yielding objective $\approx 0$. This is a closed-system accounting requirement: every unit in the system must originate from a production decision.
    
    \item \textbf{Aging direction convention ($k$ decrements toward expiration).}
    Inventory ages by decrementing $k$: freshly produced units enter at $k = \text{SL}_p$ and expire when they reach $k = 1$. The reverse convention (incrementing $k$) would place new arrivals at $k = 1$ and ``age'' them toward $k = \text{SL}_p + 1$, which does not exist, so no unit would ever expire. The decrementing convention gives every unit a finite shelf life, after which it has either been sold or become waste.
    
    \item \textbf{Guarded temporal boundaries.}
    Aging constraints are defined only for $t \in \{1, \dots, T{-}1\}$, and lead-time arrivals $A_{p,l,t}$ default to zero when $t - \text{LT}_p < 1$. These guards prevent out-of-range indexing (e.g., constraints referencing $t = T{+}1$ or $Q_{p,l,0}$), which would cause infeasibility or runtime errors rather than silent modeling mistakes.
    
    \item \textbf{Directional substitution edges.}
    Each edge $(p_{\text{from}}, p_{\text{to}}) \in \mathcal{E}^{\text{sub}}$ is \emph{asymmetric}: $p_{\text{from}}$'s demand can be served by $p_{\text{to}}$'s inventory, but not vice versa. This models upward substitution (e.g., premium stock serving basic demand). The prompt text and JSON schema both encode the direction explicitly via ordered pairs \texttt{[p\_from, p\_to]} to minimize ambiguity. Pilot experiments showed that reversing the substitution direction is one of the most common modeling mistakes (${\sim}35\%$ error rate across four frontier LLMs).
    
    \item \textbf{Holding cost restricted to $k \ge 2$.}
    Units in the oldest bucket ($k = 1$) that are unsold incur waste cost $c^{\text{waste}}_p$. Applying holding cost $c^{\text{inv}}_p$ to this same bucket would double-count the penalty. The formulation therefore charges holding cost only on end-of-period inventory with $k \ge 2$, which keeps the waste and holding cost terms separate.
\end{enumerate}

\section{Scenario Family Design}
\label{app:families}

RetailOpt-190 is constructed around four design principles.
\textbf{(1)~Structural coverage:} each instance activates at least one
core retail mechanism (perishability, capacity coupling, substitution,
discrete procurement, or network flows), so the benchmark covers a range of
formulation challenges.
\textbf{(2)~Compositionality:} difficulty arises from mechanism
\emph{interactions}, not linguistic complexity; a perishability-only
instance is straightforward, but perishability coupled with capacity
limits and substitution reveals whether models correctly propagate
constraints.
\textbf{(3)~Diagnosability:} each scenario preserves recognizable module
signatures, enabling precise failure localization through objective-value
comparison and error analysis.
\textbf{(4)~Feasibility stress:} several instances are designed so that
common modeling errors (e.g., missing inventory balance) lead to
infeasibility, providing diagnostic signals beyond objective comparison.

\subsection{Family--Mechanism Design Matrix}

The benchmark comprises 38 archetypes across 8 families
(Table~\ref{tab:family_design}).

\begin{table}[ht]
\centering
\small
\caption{Compositional design matrix. Each family tests specific mechanism interactions. \ding{51} = structurally active (affects optimal solution); \textcolor{gray}{$\circ$} = present but not binding in default data.}
\label{tab:family_design}
\footnotesize
\setlength{\tabcolsep}{3pt}
\begin{tabular}{@{}llccccccc@{}}
\toprule
\textbf{ID} & \textbf{Family} & \textbf{\#Arc.} & \textbf{Perish.} & \textbf{Capacity} & \textbf{Subst.} & \textbf{Discrete} & \textbf{Network} & \textbf{Primary Test} \\
\midrule
F1 & Core Operations         & 4 & \ding{51} & \textcolor{gray}{$\circ$} & \textcolor{gray}{$\circ$} & -- & -- & Aging dynamics \\
F2 & Assortment              & 6 & \ding{51} & \ding{51} & \ding{51} & \textcolor{gray}{$\circ$} & -- & Substitution $\times$ capacity \\
F3 & Resources               & 4 & \ding{51} & \ding{51} & \textcolor{gray}{$\circ$} & -- & -- & Multi-resource coupling \\
F4 & Dynamics                & 6 & \ding{51} & \ding{51} & \textcolor{gray}{$\circ$} & -- & -- & Temporal propagation \\
\midrule
F5 & Feasibility Stress      & 4 & \ding{51} & \ding{51} & -- & -- & -- & Stress robustness \\
F6 & Discrete Logistics      & 4 & \ding{51} & \textcolor{gray}{$\circ$} & \textcolor{gray}{$\circ$} & \ding{51} & -- & Integer constraints \\
F7 & Network \& Multi-Echelon& 6 & \ding{51} & \ding{51} & \textcolor{gray}{$\circ$} & \textcolor{gray}{$\circ$} & \ding{51} & Multi-location flow \\
F8 & Omni-channel            & 4 & \ding{51} & \ding{51} & \textcolor{gray}{$\circ$} & -- & \textcolor{gray}{$\circ$} & Returns $\times$ labor \\
\bottomrule
\end{tabular}
\end{table}

\paragraph{Difficulty Progression.}
Families F1--F4 test mechanisms with limited interaction; a model may pass by correctly implementing each module independently.
Families F5--F8 require \emph{compositional reasoning}: constraints must jointly bind, and errors in one module propagate to others.
For example, F7 (Network) requires correct capacity constraints \emph{and} transshipment flows \emph{and} budget limits simultaneously.
F5 (Feasibility Stress) deliberately combines stressors so that a single modeling omission (e.g., missing lost-sales slack) renders the problem infeasible, while the ground-truth formulation, which includes all necessary slack variables, remains feasible at a very high cost. As a result, an infeasible status returned by an LLM's code on an F5 instance indicates a structural modeling error rather than a data issue.

\subsection{Archetype Specifications}
\label{app:archetype_specs}

Each archetype specifies: (1)~active modules, (2)~parameter modifications relative to the base scenario, and (3)~expected binding patterns.
Table~\ref{tab:archetypes_full} provides the complete list of all 38 archetypes with their structural modifications.

\paragraph{Naming Convention.}
Each archetype is identified by \texttt{retail\_\{family\}\_\{descriptor\}} in the codebase (e.g., \texttt{retail\_f2\_no\_substitution}).
Table~\ref{tab:archetypes_full} omits the \texttt{retail\_} prefix for brevity; the suffix directly corresponds to the archetype key in \texttt{archetypes.yaml} and the generator function name in \texttt{retail\_benchmark\_generator.py}.
Instance files are named \texttt{retail\_\{family\}\_\{descriptor\}\_v\{0--4\}.json}.

\paragraph{Reading the Table.}
The ``Modification'' column describes only the parameters that change relative to the base scenario (\S\ref{app:base_scenario}); all unmentioned fields retain their base values.
Numeric values (e.g., ``$\times 0.3$'') are multiplicative factors applied in the generator code.
Where specific cost values or per-SKU arrays are listed, they correspond to the order \texttt{[SKU\_Basic, SKU\_Premium, SKU\_ShortLife]}.

\begin{table}[ht]
\centering
\small
\caption{Complete archetype specifications. ``Modification'' describes the parameter change relative to the base scenario (\texttt{get\_base\_scenario()} in the generator). All 38 archetypes share the same JSON schema and are solved by the same universal MILP.}
\label{tab:archetypes_full}
\setlength{\tabcolsep}{3pt}
\begin{tabular}{@{}llp{7.8cm}@{}}
\toprule
\textbf{Family} & \textbf{Archetype ID} & \textbf{Modification from Base Scenario} \\
\midrule
\multirow{4}{*}{F1} 
  & \texttt{f1\_base}         & Baseline: 20 periods, 3 SKUs, 5 DCs, seasonal Gaussian demand \\
  & \texttt{f1\_high\_waste}  & Waste cost $\times 20$ for all products \\
  & \texttt{f1\_jit\_logic}   & Holding cost $\times 20$ for all products \\
  & \texttt{f1\_52\_weeks}    & Horizon extended to $T{=}52$ (demand/capacity arrays tiled from base) \\
\midrule
\multirow{6}{*}{F2}
  & \texttt{f2\_no\_substitution}  & Substitution edges removed ($\mathcal{E}^{\text{sub}} = \emptyset$) \\
  & \texttt{f2\_circular\_sub}     & Circular substitution ring: Basic$\to$Premium$\to$ShortLife$\to$Basic \\
  & \texttt{f2\_cannibalization}   & Basic demand $\times 2$, Basic lost-sales penalty reduced to \$5, storage $\times 0.5$ \\
  & \texttt{f2\_ultra\_fresh}      & Shelf life reduced to $\{2,2,1\}$ periods \\
  & \texttt{f2\_price\_band\_tight}& Premium purchasing cost $\times 0.8$, Basic $\times 1.1$, Premium lost-sales $\times 2$ \\
  & \texttt{f2\_promo\_budget}     & Last 4 periods: Basic/ShortLife demand $\times 2$; budget = \$15{,}000/period \\
\midrule
\multirow{4}{*}{F3}
  & \texttt{f3\_storage\_bottleneck}    & Cold capacity $\times 0.3$ at all locations \\
  & \texttt{f3\_volumetric\_constraint} & Premium cold usage increased to 15.0 (vs.\ 3.0 baseline) \\
  & \texttt{f3\_supply\_bottleneck}     & Production capacity $\times 0.3$; storage set to 999{,}999 \\
  & \texttt{f3\_unbalanced\_network}    & DC1 gets 96\% of total storage; others get 1\% each \\
\midrule
\multirow{6}{*}{F4}
  & \texttt{f4\_early\_stockout}   & Production = 0 in periods 1--5 \\
  & \texttt{f4\_peak\_failure}     & Production = 0 in periods 9--12 (peak demand window) \\
  & \texttt{f4\_demand\_surge}     & Period-15 demand $\times 4$ (single-period spike) \\
  & \texttt{f4\_quality\_hold}     & Basic production = 0 from period 11 onward \\
  & \texttt{f4\_robust\_variance}  & Alternating demand $\times 1.5$/$\times 0.7$; lost-sales $\times 2.5$ \\
  & \texttt{f4\_supply\_risk}      & Mid-horizon production $\times 0.4$ for 4 periods; waste cost $\times 3$ \\
\midrule
\multirow{4}{*}{F5}
  & \texttt{f5\_impossible\_demand}    & All demand $\times 5$ (far exceeds production capacity) \\
  & \texttt{f5\_strict\_service\_trap} & Storage $\times 0.1$ (near-zero capacity) \\
  & \texttt{f5\_storage\_overflow}     & Cold capacity = 0.5 at all locations \\
  & \texttt{f5\_ultimate\_stress}      & Composite: storage $\times 0.3$ + peak failure (periods 9--12) + no substitution \\
\midrule
\multirow{4}{*}{F6}
  & \texttt{f6\_lead\_time}          & Lead times = $\{3, 4, 2\}$ periods per SKU \\
  & \texttt{f6\_moq\_binary}         & MOQ = 300 units (global) \\
  & \texttt{f6\_fixed\_order\_cost}   & Fixed ordering cost = \$5{,}000 per order \\
  & \texttt{f6\_pack\_size\_integer}  & Pack size = 100 units \\
\midrule
\multirow{6}{*}{F7}
  & \texttt{f7\_transshipment}       & Full bidirectional transshipment among all 5 DCs \\
  & \texttt{f7\_hub\_and\_spoke}     & Hub (DC1: 50{,}000 cap) $\to$ 4 spokes (500 cap each) \\
  & \texttt{f7\_budget\_limit}       & Per-period budget = \$10{,}000 \\
  & \texttt{f7\_multi\_sourcing}     & Heterogeneous lead times $\{5,0,1\}$; holding costs Basic=0.5, Premium=10.0 \\
  & \texttt{f7\_multiechelon\_chain} & 3-tier: Plant $\to$ 2 DCs $\to$ 3 Stores (6 locations; Plant/DC demand = 0) \\
  & \texttt{f7\_ring\_routing}       & Circular transshipment ring among 5 DCs; storage $\times 0.8$ \\
\midrule
\multirow{4}{*}{F8}
  & \texttt{f8\_reverse\_logistics}   & Return rates $\{0.20, 0.10, 0.05\}$ per SKU \\
  & \texttt{f8\_labor\_constraint}    & Labor cap = 200/period; usage = $\{0.1, 0.2, 0.1\}$ per SKU \\
  & \texttt{f8\_ship\_from\_store}    & Storage $\times 5$; labor cap = 500; high per-unit labor usage $\{0.5, 0.8, 0.6\}$\\
  & \texttt{f8\_sustainability}       & Global waste cap $\le$ 2\% of total demand \\
\bottomrule
\end{tabular}
\end{table}

\subsection{Instance Generation via Controlled Perturbation}
\label{app:perturbation}

Each archetype is expanded into 5 numerical variants (v0--v4), yielding $38 \times 5 = 190$ instances total. Variant v0 uses unperturbed parameters; variants v1--v4 apply controlled stochastic perturbation with intensity $\alpha = 0.15$.

\paragraph{Perturbation Procedure.}
For each variant $v \in \{1,2,3,4\}$ and archetype with base name $n$:
\begin{enumerate}[nosep]
    \item Compute a deterministic seed: $\texttt{seed} = \texttt{uint32\_le}\!\big(\texttt{SHA256}(n\,\texttt{"|\!"|}\,v)\,[{:}4]\big)$, where the input string is \texttt{"\{name\}|\{v\}"} and \texttt{uint32\_le} reads the first 4 bytes of the digest as an unsigned 32-bit little-endian integer.
    \item Initialize a NumPy random generator: $\texttt{rng} = \texttt{default\_rng}(\texttt{seed})$
    \item Perturb demand curves: $\tilde{d}_{p,t} = \lfloor d_{p,t} \cdot U(1{-}\alpha, 1{+}\alpha) \rfloor$ for each product and period
    \item Perturb storage capacities: $\tilde{C}_l = C_l \cdot U(1{-}\alpha, 1{+}\alpha)$ for each location
\end{enumerate}
where $U(a,b)$ denotes independent draws from $\text{Uniform}(a,b)$ and $\alpha = 0.15$.

\paragraph{Rationale.}
The perturbation targets the two parameter groups most likely to change which constraints bind, demand volumes and storage capacities, and keeps the structure fixed (shelf life, network topology, cost ratios). The deterministic hashing makes the instances reproducible across platforms without a fixed global seed.

\subsection{Base Scenario Specification}
\label{app:base_scenario}

All 38 archetypes derive from a single base scenario via modular parameter overrides. The base scenario is generated by \texttt{get\_base\_scenario()} in the benchmark generator and describes a moderately complex retail setting: three product tiers with heterogeneous shelf lives, five distribution centers with non-uniform demand shares, and a seasonal demand pattern that peaks mid-horizon. Costs are calibrated so that lost sales dominate waste in the objective, and the optimal plan therefore trades off overstocking (waste) against understocking (lost-sales penalty). Table~\ref{tab:base_params} summarizes the base configuration.

\begin{table}[ht]
\centering
\small
\caption{Base scenario parameters shared across all archetypes (before archetype-specific modifications).}
\label{tab:base_params}
\begin{tabular}{@{}llp{5.4cm}@{}}
\toprule
\textbf{Parameter} & \textbf{Value} & \textbf{Description} \\
\midrule
$T$ & 20 & Planning periods \\
$|\mathcal{P}|$ & 3 & SKU\_Basic, SKU\_Premium, SKU\_ShortLife \\
$|\mathcal{L}|$ & 5 & DC1 through DC5 \\
Shelf life & $\{10, 8, 4\}$ & Periods remaining at production \\
Lead time & $\{0, 0, 0\}$ & Same-period arrival (default) \\
\midrule
Demand curve & Gaussian: $\lfloor 1000 \exp\!\big(\!-\frac{(t-10)^2}{18}\big) + 300 \rfloor$ & Seasonal peak at mid-horizon ($t$ is 0-indexed); \newline SKU\_Premium $\times 0.5$, SKU\_ShortLife $\times 0.4$; \newline values cast to \texttt{int} in generator \\
Demand share & $\{0.25, 0.20, 0.20, 0.20, 0.15\}$ & Allocated to DC1--DC5 \\
\midrule
Production cap & $\{800, 400, 500\}$/period & Per-SKU, constant across periods \\
Storage cap & $\{4000, 3500, 3000, 3000, 2500\}$ & Per-DC cold capacity \\
Cold usage & $\{1.0, 3.0, 1.2\}$ & Volume units per product unit \\
\midrule
$c^{\text{buy}}$ & $\{10, 20, 15\}$ & Purchasing cost per unit \\
$c^{\text{inv}}$ & $\{1.0, 1.5, 1.0\}$ & Holding cost per unit per period \\
$c^{\text{waste}}$ & $\{2.0, 3.0, 2.0\}$ & Waste cost per unit expired \\
$c^{\text{ls}}$ & $\{50, 80, 40\}$ & Lost-sales penalty per unit \\
$c^{\text{fix}}$ & $0$ & No fixed ordering cost \\
$c^{\text{tr}}$ & $0.5$ & Transshipment cost per unit \\
\midrule
Substitution & Basic $\to$ Premium & Upward: Premium can serve Basic's demand \\
Transshipment & $\emptyset$ & No inter-location movement (default) \\
MOQ, pack size & $0, 1$ & Continuous ordering (default) \\
\bottomrule
\end{tabular}
\end{table}

\section{Data Formats and Access}
\label{app:data_format}

\subsection{Prompt Format Comparison}

RetailOpt-190 provides two prompt formats per scenario to support different evaluation paradigms (Table~\ref{tab:prompt_formats}).

\begin{table}[ht]
\centering
\small
\caption{Two prompt formats for different evaluation scenarios. The \emph{same semantic information} is conveyed in both formats; only the data delivery method differs.}
\label{tab:prompt_formats}
\begin{tabularx}{\textwidth}{@{}l>{\raggedright\arraybackslash}p{2.0cm}>{\raggedright\arraybackslash}X>{\raggedright\arraybackslash}p{2.1cm}l@{}}
\toprule
\textbf{Format} & \textbf{Data Location} & \textbf{Content} & \textbf{Use Case} & \textbf{File Suffix} \\
\midrule
Schema-based & External (runtime) & Narrative + JSON schema (types only) & Scalability, production & \texttt{.scenario.txt} \\
Data-embedded & In prompt          & Narrative + full JSON data           & Benchmark comparison   & \texttt{.full.txt} \\
\bottomrule
\end{tabularx}
\end{table}

\paragraph{Default Reporting Format.}
We report all results using the \emph{data-embedded} format (\texttt{.full.txt}) for fair comparison with NL4Opt~\citep{ramamonjison2023nl4opt}, MAMO~\citep{huang2024mamo}, and IndustryOR~\citep{huang2025orlm}, which embed data directly in prompts. The schema-based format is also available for evaluating LLM data-access capabilities in production-like settings.

\paragraph{Two Data Delivery Mechanisms.}
The schema-based and data-embedded formats convey identical semantic information but differ in how instance data reaches the LLM-generated code:
\begin{itemize}[nosep]
    \item \textbf{Schema-based (\texttt{.scenario.txt}):} The prompt contains only a \emph{type-level} JSON schema (e.g., \texttt{"periods": int}, \texttt{"demand\_curve": \{p: [float]\}}) and a \texttt{[DATA ACCESS]} section instructing the LLM that ``The variable \texttt{data} is pre-loaded. Do NOT use file I/O.'' The actual numerical JSON is injected at runtime as a Python dictionary. The LLM must write code that reads from the \texttt{data} dict using correct key names, nested access patterns, and 0-to-1 index conversion.
    \item \textbf{Data-embedded (\texttt{.full.txt}):} The prompt contains the \emph{complete numerical JSON} inline within a \texttt{[DATA]} section. There is no \texttt{[DATA SCHEMA]} or \texttt{[DATA ACCESS]} section. The LLM must embed the JSON as a string literal and parse it with \texttt{json.loads()}, or hardcode the values directly.
\end{itemize}
\noindent This design isolates the effect of data access complexity: the schema-based format tests whether LLMs can correctly read nested JSON structures from type signatures alone, while the data-embedded format tests whether they can extract structure from raw numerical data. Section~\ref{app:worked_example} provides a concrete side-by-side comparison.

\paragraph{Prompt Architecture.}
The prompt is \emph{moderately scaffolded}. It provides the business narrative, the data, and a set of \emph{structural cues}: variable indexing conventions, the perishable inflow accounting equation aligned with the reference MILP (per-location order quantity $Q_{p,l,t}$, lead-time arrival, transshipment net flow, and returns), and the aggregate production constraint. It does \textbf{not} provide a complete formulation, an objective function specification, the full constraint set, or warnings about common errors.
The cues are necessary because fully unscaffolded prompts produce near-zero accuracy on compositional retail problems for all tested models. With moderate scaffolding, the benchmark tests whether an LLM can compose the remaining optimization structure (objective, capacity/demand constraints, integrality reasoning, multi-period coupling) from the natural-language description.

\begin{table}[ht]
\centering
\small
\caption{Information provided in baseline prompts vs.\ left for the LLM to derive.}
\label{tab:prompt_content}
\begin{tabular}{@{}ll@{}}
\toprule
\textbf{Provided in Prompt} & \textbf{Left for LLM to Derive} \\
\midrule
Business narrative with structure cues & Complete formulation (variables, objective, constraints) \\
Data schema (field names, types) & Capacity, demand, and budget constraints \\
Data access patterns (indexing, nesting) & Substitution and inventory dynamics constraints \\
Output format specification (GurobiPy) & Boundary conditions / edge cases \\
Inventory inflow accounting (URS-aligned) & Common error warnings \\
\bottomrule
\end{tabular}
\end{table}

\subsection{JSON Schema}
\label{app:json_schema}

All 190 instances share a universal JSON schema, ensuring that a single solver implementation can process every archetype without format-specific logic. The schema is organized into six groups: structural dimensions (periods, products, locations), perishability parameters (shelf life, lead time), demand specification (curve and share), capacity limits (production, storage, labor), cost coefficients, and network topology. Table~\ref{tab:schema_fields} provides a complete field reference with types, default values from the base scenario, and semantic descriptions. Fields under \texttt{constraints} and \texttt{network} are nested objects; safe access via \texttt{data.get()} with defaults is required to avoid \texttt{KeyError} on archetypes where these fields are absent (see Pattern~3 in \S\ref{app:access_patterns}).

\begin{table}[ht]
\centering
\small
\caption{Complete JSON schema field reference. Types: \texttt{int} = integer, \texttt{float} = real, \texttt{str} = string, \texttt{[T]} = array of type T, \texttt{\{K:V\}} = dict, \texttt{null} = absent/inactive.}
\label{tab:schema_fields}
\footnotesize
\setlength{\tabcolsep}{3pt}
\begin{tabularx}{\textwidth}{@{}lll>{\raggedright\arraybackslash}X@{}}
\toprule
\textbf{Field} & \textbf{Type} & \textbf{Example} & \textbf{Semantics} \\
\midrule
\texttt{periods} & \texttt{int} & 20 & Number of planning periods $T$ \\
\texttt{products} & \texttt{[str]} & \texttt{["SKU\_Basic",...]} & Product identifiers \\
\texttt{locations} & \texttt{[str]} & \texttt{["DC1",...]} & Location identifiers \\
\texttt{shelf\_life} & \texttt{\{str:int\}} & \texttt{\{"SKU\_Basic":10\}} & Remaining-life capacity per SKU \\
\texttt{lead\_time} & \texttt{\{str:int\}} & \texttt{\{"SKU\_Basic":0\}} & Order-to-arrival delay (periods) \\
\midrule
\texttt{demand\_curve} & \texttt{\{str:[float]\}} & \texttt{[303,311,...]} & Aggregate demand per SKU per period (0-indexed) \\
\texttt{demand\_share} & \texttt{\{str:float\}} & \texttt{\{"DC1":0.25\}} & Location's fraction of aggregate demand \\
\texttt{production\_cap} & \texttt{\{str:[float]\}} & \texttt{[800,...]} & Max production per SKU per period (0-indexed) \\
\texttt{cold\_capacity} & \texttt{\{str:float\}} & \texttt{\{"DC1":4000\}} & Storage capacity per location \\
\texttt{cold\_usage} & \texttt{\{str:float\}} & \texttt{\{"SKU\_Basic":1.0\}} & Volume units per product unit \\
\midrule
\texttt{labor\_cap} & \texttt{\{str:[float]\}} & \texttt{[99999,...]} & Labor hours per location per period \\
\texttt{labor\_usage} & \texttt{\{str:float\}} & \texttt{\{"SKU\_Basic":0.0\}} & Labor hours per unit sold \\
\texttt{return\_rate} & \texttt{\{str:float\}} & \texttt{\{"SKU\_Basic":0.0\}} & Fraction of sales returned next period \\
\midrule
\texttt{costs.purchasing} & \texttt{\{str:float\}} & \texttt{\{"SKU\_Basic":10\}} & Per-unit buying cost \\
\texttt{costs.inventory} & \texttt{\{str:float\}} & \texttt{\{"SKU\_Basic":1.0\}} & Per-unit holding cost per period \\
\texttt{costs.waste} & \texttt{\{str:float\}} & \texttt{\{"SKU\_Basic":2.0\}} & Per-unit expiration penalty \\
\texttt{costs.lost\_sales} & \texttt{\{str:float\}} & \texttt{\{"SKU\_Basic":50\}} & Per-unit lost-demand penalty \\
\texttt{costs.fixed\_order} & \texttt{float} & 0.0 & Fixed cost per order placed \\
\texttt{costs.transshipment} & \texttt{float} & 0.5 & Per-unit transshipment cost \\
\midrule
\texttt{constraints.moq} & \texttt{float} & 0 & Minimum order quantity (0 = inactive) \\
\texttt{constraints.pack\_size} & \texttt{int} & 1 & Pack size (1 = continuous) \\
\texttt{constraints.budget\_per\_period} & \texttt{float|null} & \texttt{null} & Per-period purchasing budget \\
\texttt{constraints.waste\_limit\_pct} & \texttt{float|null} & \texttt{null} & Max waste as fraction of total demand \\
\midrule
\texttt{network.sub\_edges} & \texttt{[[str,str]]} & \texttt{[["Basic","Premium"]]} & Directed substitution: from's demand served by to's inventory \\
\texttt{network.trans\_edges} & \texttt{[[str,str]]} & \texttt{[]} & Directed transshipment arcs between locations \\
\bottomrule
\end{tabularx}
\end{table}

\subsection{Critical Data Access Patterns}
\label{app:access_patterns}

Three data access patterns cause frequent LLM errors:

\paragraph{Pattern 1: Demand Indexing Mismatch.}
The JSON arrays are 0-indexed, but the optimization model uses 1-indexed periods:
\begin{equation}
d_{p,l,t} = \texttt{demand\_curve}[p][t{-}1] \times \texttt{demand\_share}[l], \qquad t \in \{1,\dots,T\}
\end{equation}
LLMs frequently access \texttt{demand\_curve[p][t]} directly, causing an off-by-one error that shifts the entire demand profile by one period.

\paragraph{Pattern 2: Substitution Edge Semantics.}
The JSON encodes \texttt{sub\_edges} as \texttt{[p\_from, p\_to]}, meaning \emph{$p_{\text{from}}$'s demand can be served by $p_{\text{to}}$'s inventory}. This is ``upward substitution'': a premium product serves a basic product's excess demand.
In the reference solver, this translates to:
\begin{itemize}[nosep]
    \item $\mathcal{N}^{\text{out}}_p$: products whose inventory can serve $p$'s demand ($p$ exports demand to them)
    \item $\mathcal{N}^{\text{in}}_p$: products whose demand $p$'s inventory serves ($p$ receives their demand)
\end{itemize}
For example, edge \texttt{["SKU\_Basic", "SKU\_Premium"]} creates $\mathcal{N}^{\text{out}}_{\text{Basic}} = \{\text{Premium}\}$ and $\mathcal{N}^{\text{in}}_{\text{Premium}} = \{\text{Basic}\}$.

\paragraph{Pattern 3: Nested Network Access.}
Network data is nested within the \texttt{network} object. Safe access requires:
\begin{lstlisting}[basicstyle=\ttfamily\scriptsize,language=Python,numbers=none]
sub_edges = [tuple(e) for e in data.get('network', {}).get('sub_edges', [])]
trans_edges = [tuple(e) for e in data.get('network', {}).get('trans_edges', [])]
\end{lstlisting}
LLMs that access \texttt{data['sub\_edges']} directly encounter a \texttt{KeyError}, converting a potential silent failure into an execution failure.

\subsection{Worked Example: \texttt{retail\_f1\_52\_weeks\_v0}}
\label{app:worked_example}

We illustrate the complete data pipeline using the \texttt{retail\_f1\_52\_weeks\_v0} instance (F1 family, 52-week horizon). This archetype tiles the base 20-period demand and capacity arrays to 52 periods, testing long-horizon accumulation effects while preserving the same constraint structure.

\paragraph{Instance JSON (Truncated).}
Listing~\ref{lst:json_example} shows the JSON structure with arrays abbreviated. The full instance contains 52-element arrays for \texttt{demand\_curve}, \texttt{production\_cap}, and \texttt{labor\_cap}; all other fields are scalars or short dicts.

\newpage
\begin{lstlisting}[basicstyle=\ttfamily\scriptsize,caption={Instance JSON for \texttt{retail\_f1\_52\_weeks\_v0} (arrays truncated for space).},label=lst:json_example,numbers=none,frame=single]
{
  "name": "retail_f1_52_weeks_v0",
  "description": "Standard seasonal retail scenario.",
  "periods": 52,
  "products": ["SKU_Basic", "SKU_Premium", "SKU_ShortLife"],
  "locations": ["DC1", "DC2", "DC3", "DC4", "DC5"],
  "shelf_life": {"SKU_Basic": 10, "SKU_Premium": 8, "SKU_ShortLife": 4},
  "lead_time":  {"SKU_Basic": 0,  "SKU_Premium": 0, "SKU_ShortLife": 0},
  "cold_capacity": {"DC1": 4000, "DC2": 3500, "DC3": 3000,
                    "DC4": 3000, "DC5": 2500},
  "cold_usage": {"SKU_Basic": 1.0, "SKU_Premium": 3.0,
                 "SKU_ShortLife": 1.2},
  "production_cap": {
    "SKU_Basic":     [800, 800, ..., 800],       // 52 elements
    "SKU_Premium":   [400, 400, ..., 400],
    "SKU_ShortLife": [500, 500, ..., 500]
  },
  "labor_cap": {"DC1": [99999.0, ...], ...},     // 5 x 52
  "labor_usage":  {"SKU_Basic": 0.0, ...},
  "return_rate":  {"SKU_Basic": 0.0, ...},
  "demand_curve": {
    "SKU_Basic":     [303, 311, 328, ..., 1300, 1245],
    "SKU_Premium":   [151, 155, 164, ...,  650,  622],
    "SKU_ShortLife": [121, 124, 131, ...,  520,  498]
  },
  "demand_share": {"DC1": 0.25, "DC2": 0.2, "DC3": 0.2,
                   "DC4": 0.2,  "DC5": 0.15},
  "costs": {
    "lost_sales": {"SKU_Basic":50,  "SKU_Premium":80,  "SKU_ShortLife":40},
    "inventory":  {"SKU_Basic":1.0, "SKU_Premium":1.5, "SKU_ShortLife":1.0},
    "waste":      {"SKU_Basic":2.0, "SKU_Premium":3.0, "SKU_ShortLife":2.0},
    "fixed_order": 0.0, "transshipment": 0.5,
    "purchasing": {"SKU_Basic":10, "SKU_Premium":20, "SKU_ShortLife":15}
  },
  "constraints": {"moq": 0, "pack_size": 1,
                  "budget_per_period": null, "waste_limit_pct": null},
  "network": {
    "sub_edges":   [["SKU_Basic", "SKU_Premium"]],
    "trans_edges": []
  }
}
\end{lstlisting}

\paragraph{Schema-Based Prompt (\texttt{.scenario.txt}).}
Listing~\ref{lst:prompt_schema} shows the schema-based prompt structure. The prompt provides the business narrative with embedded structure cues (perishability equations, substitution semantics), a type-level JSON schema, data access instructions, and the required output format. It does \emph{not} include the actual numerical data; at evaluation time, the JSON is loaded externally and made available as the \texttt{data} variable.

\begin{lstlisting}[basicstyle=\ttfamily\scriptsize,caption={Schema-based prompt (\texttt{.scenario.txt}, abbreviated). The full prompt is 112 lines; sections marked \texttt{...} indicate omitted material of similar structure.},label=lst:prompt_schema,numbers=none,frame=single,breaklines=true]
[SCENARIO]
Family: F1 (Core Operations)
Archetype: retail_f1_52_weeks
Scenario ID: retail_f1_52_weeks_v0

[BUSINESS DESCRIPTION]
Business narrative:
The retailer plans inventory over a full year of weekly decisions,
represented by fifty-two time periods. Exogenous seasonal demand,
local inventories at each distribution center, and per-product
production capacities work as in the core operations baseline.
Customers are never backordered; unmet demand in a week is
immediately lost and penalized. ...

Structure cues:
- ...
- Shelf life: Each product has a shelf life in periods.
  Inventory must be tracked by REMAINING LIFE.

  VARIABLE DEFINITION: I[p,l,t,r] = inventory at START of period t
  with r periods remaining.
  Convention: r=1 is OLDEST (sell first FIFO),
              r=shelf_life[p] is FRESHEST.

  KEY EQUATIONS - implement these conventions consistently with
  the JSON data:

  (1) Fresh inflow:
      I[p,l,t,SL] = Q[p,l,t-LT[p]] + transshipment_net[p,l,t]
                   + returns[p,l,t]
      - Q[p,l,t] is the per-LOCATION order quantity
      - LT[p] = lead_time[p]; treat Q[p,l,s] as 0 for s < 1
      - Aggregate production constraint:
          sum over locations of Q[p,l,t] <= production_cap[p][t]
      - transshipment_net = inbound - outbound flows on trans_edges
        (zero when network.trans_edges is empty)
      - returns = return_rate[p] * sum_a sales[p,l,t-1,a]
        when t > 1, else 0
      - This is ONLY the fresh-inflow channel; do NOT subtract
        sales here.
  (2) Aging: I[p,l,t+1,r] = I[p,l,t,r+1] - sales[p,l,t,r+1]
      for r=1..SL-1
  (3) Waste: W[p,l,t] = I[p,l,t,1] - sales[p,l,t,1]
  (4) Sales availability: sales[p,l,t,r] <= I[p,l,t,r]
  (5) Inventory holding cost: charged on
      (I[p,l,t,r] - sales[p,l,t,r]) for r >= 2

- Substitution: Edge [p_from, p_to] means p_from's demand can
  be served by p_to's inventory.
  Variable sub[p_from, p_to, l, t] = units of p_from's demand
  fulfilled by p_to.

  Demand fulfillment equation:
  - For p_from: total_sales[p_from] + sub[p_from, p_to]
                + L[p_from] = demand[p_from]
  - For p_to:   total_sales[p_to] - sub[p_from, p_to]
                + L[p_to]   = demand[p_to]

- No transshipment and zero lead times in this scenario.
- The objective is to minimize total cost over the entire year,
  aggregating purchasing, holding, waste, and lost sales costs across all weeks.

[DATA SCHEMA]
{
  "periods": int,       "products": [str, ...],
  "shelf_life": {p: int},  "lead_time": {p: int},
  "demand_curve": {p: [float, ...]},
  "demand_share": {l: float},
  "production_cap": {p: [float, ...]},
  "cold_capacity": {l: float},  "cold_usage": {p: float},
  "costs": { "purchasing": {p: float}, "inventory": {p: float},
             "waste": {p: float}, "lost_sales": {p: float},
             "fixed_order": float, "transshipment": float },
  "constraints": { "moq": float, "pack_size": int,
                   "budget_per_period": float|null,
                   "waste_limit_pct": float|null },
  "network": { "sub_edges": [[p_from, p_to], ...],
               "trans_edges": [[l_from, l_to], ...] }
}

[DATA ACCESS]
- The variable `data` is pre-loaded. Do NOT use file I/O.
- Lists are 0-indexed (period t in model uses index [t-1] in
  data arrays)

CRITICAL - Network edges require tuple conversion for Gurobi:
  sub_edges = [tuple(e) for e in
               data.get('network', {}).get('sub_edges', [])]
  trans_edges = [tuple(e) for e in
               data.get('network', {}).get('trans_edges', [])]

[OUTPUT FORMAT]
- Import: import gurobipy as gp; from gurobipy import GRB
- Set Gurobi params: m.Params.OutputFlag = 0;
  m.Params.Threads = 1; m.Params.Seed = 0
- Print at end:
  print(f"status: {{m.Status}}")
  if m.Status == 2:
      print(f"objective: {{m.ObjVal}}")
- Output ONLY executable Python code. No markdown, no explanations.

[TASK]
Write a GurobiPy script that models and solves this optimization
problem.
\end{lstlisting}

\paragraph{Data-Embedded Prompt (\texttt{.full.txt}).}
The data-embedded format delivers the \emph{same business narrative} as the schema-based prompt but changes \textbf{how data reaches the LLM-generated code}.
In the schema-based format, the prompt provides only field types (e.g., \texttt{"periods": int}) and instructs the LLM to read from a pre-loaded \texttt{data} variable at runtime.
In the data-embedded format, the prompt contains the \emph{complete numerical JSON} inline, and the LLM must parse it with \texttt{json.loads()}.
The key structural differences are:

\begin{enumerate}[nosep]
    \item \textbf{Schema-based} has \texttt{[DATA SCHEMA]} (types only) $+$ \texttt{[DATA ACCESS]} (tells LLM: ``\texttt{data} is pre-loaded, do NOT use file I/O'').
    \item \textbf{Data-embedded} replaces both with \texttt{[DATA]} containing the full JSON (694 lines for this instance). No \texttt{[DATA SCHEMA]} or \texttt{[DATA ACCESS]} sections appear.
    \item The \texttt{[OUTPUT FORMAT]} in the data-embedded format also requires \texttt{import json}, and the \texttt{[TASK]} explicitly instructs ``parse the JSON data above (use \texttt{json.loads} on the string)''.
\end{enumerate}

\noindent Listing~\ref{lst:prompt_full} shows only the sections that structurally differ from the schema-based prompt (Listing~\ref{lst:prompt_schema}). This is the default reporting format, as it enables self-contained evaluation without external file access, which matches the setup of NL4Opt~\citep{ramamonjison2023nl4opt}, MAMO~\citep{huang2024mamo}, and IndustryOR~\citep{huang2025orlm}.

\begin{lstlisting}[basicstyle=\ttfamily\scriptsize,caption={Data-embedded prompt (\texttt{.full.txt}): sections that differ from the schema-based format. The \texttt{[BUSINESS DESCRIPTION]} is identical and omitted here.},label=lst:prompt_full,numbers=none,frame=single,breaklines=true]
[SCENARIO]
Family: F1 (Core Operations)
Archetype: retail_f1_52_weeks
Scenario ID: retail_f1_52_weeks_v0

[BUSINESS DESCRIPTION]
...  (identical to .scenario.txt -- same narrative and equations)

                     *** NO [DATA SCHEMA] section ***
                     *** NO [DATA ACCESS] section ***

[DATA]
The following JSON contains all instance data. Parse it directly
in your code.

```json
{
  "name": "retail_f1_52_weeks_v0",
  "description": "Standard seasonal retail scenario.",
  "periods": 52,
  "products": ["SKU_Basic", "SKU_Premium", "SKU_ShortLife"],
  "locations": ["DC1", "DC2", "DC3", "DC4", "DC5"],
  "shelf_life": {"SKU_Basic": 10, "SKU_Premium": 8, ...},
  "demand_curve": {
    "SKU_Basic": [303, 311, 328, 365, 435, 549, 711, 906,
                  1100, 1245, 1300, 1245, 1100, 906, 711,
                  549, 435, 365, 328, 311,          // period 1-20
                  303, 311, 328, ..., 1300, 1245],   // tiled to 52
    "SKU_Premium": [151, 155, 164, ..., 650, 622],
    "SKU_ShortLife": [121, 124, 131, ..., 520, 498]
  },
  "production_cap": {"SKU_Basic": [800, 800, ..., 800], ...},
  "costs": {
    "lost_sales": {"SKU_Basic": 50.0, ...},
    "inventory": {"SKU_Basic": 1.0, ...},
    ...
  },
  "constraints": {"moq": 0, "pack_size": 1,
                  "budget_per_period": null,
                  "waste_limit_pct": null},
  "network": {"sub_edges": [["SKU_Basic","SKU_Premium"]],
              "trans_edges": []}
}
```

[OUTPUT FORMAT]
- Import: import gurobipy as gp; from gurobipy import GRB;
  import json
- Set Gurobi params: m.Params.OutputFlag = 0;
  m.Params.Threads = 1; m.Params.Seed = 0
- Print at end:
  print(f"status: {{m.Status}}")
  if m.Status == 2:
      print(f"objective: {{m.ObjVal}}")
- Output ONLY executable Python code. No markdown, no explanations.

[TASK]
Write a GurobiPy script that:
1. Parses the JSON data above (use json.loads on the string)
2. Models and solves the optimization problem
3. Prints status and objective value
\end{lstlisting}

\noindent Table~\ref{tab:format_contrast} summarizes the structural contrast between the two formats.

\begin{table}[ht]
\centering
\small
\caption{Section-level comparison of the two prompt formats for a single instance. \ding{51} = present, -- = absent.}
\label{tab:format_contrast}
\begin{tabular}{@{}lcc@{}}
\toprule
\textbf{Prompt Section} & \textbf{Schema-based} & \textbf{Data-embedded} \\
\midrule
\texttt{[SCENARIO]} header         & \ding{51} & \ding{51} \\
\texttt{[BUSINESS DESCRIPTION]}    & \ding{51} & \ding{51} \\
\texttt{[DATA SCHEMA]} (types)     & \ding{51} & --        \\
\texttt{[DATA ACCESS]} (runtime)   & \ding{51} & --        \\
\texttt{[DATA]} (full JSON inline) & --        & \ding{51} \\
\texttt{[OUTPUT FORMAT]}           & \ding{51} & \ding{51} \\
\texttt{[TASK]}                    & \ding{51} & \ding{51} \\
\midrule
Data delivery method  & \texttt{data} var pre-loaded & \texttt{json.loads()} in code \\
Lines (this instance) & 112       & 765       \\
\bottomrule
\end{tabular}
\end{table}

\section{Solver Configuration and Evaluation Protocol}
\label{app:solver_config}

\subsection{Ground Truth Solver Settings}

All 190 ground truth solutions are computed using the Universal Retail Solver (URS) implemented in GurobiPy. Table~\ref{tab:solver_settings} lists the Gurobi parameters used for ground truth computation. Both the URS and LLM-generated code use \texttt{Threads=1} and \texttt{Seed=0} to ensure deterministic, single-threaded reproducibility.

\begin{table}[ht]
\centering
\small
\caption{Gurobi solver parameters for ground truth computation, as set in \texttt{universal\_retail\_solver.py}.}
\label{tab:solver_settings}
\begin{tabular}{@{}lll@{}}
\toprule
\textbf{Parameter} & \textbf{Value} & \textbf{Purpose} \\
\midrule
\texttt{TimeLimit} & 60 seconds & Prevent stalling on complex MIPs (F6/F7) \\
\texttt{MIPGap} & 1\% & Tolerance for near-optimal solutions \\
\texttt{Threads} & 1 & Single-threaded for reproducibility \\
\texttt{Seed} & 0 & Fixed random seed \\
\texttt{OutputFlag} & 0 & Suppress solver output for batch processing \\
\bottomrule
\end{tabular}
\end{table}

\subsection{Accuracy Tolerances}

An instance is judged \textbf{correct} if both conditions hold:
\begin{enumerate}[nosep]
    \item \textbf{Status match}: predicted and ground-truth statuses agree (both feasible, or both infeasible).
    \item \textbf{Objective match}: for feasible instances, $|y_{\text{pred}} - y_{\text{ref}}| / |y_{\text{ref}}| < \epsilon$.
\end{enumerate}

\noindent The tolerance $\epsilon$ is family-dependent because problem structure determines the achievable precision within the 60-second time limit (Table~\ref{tab:tolerances}). Families F1--F5 and F7--F8 produce LP relaxations that are either tight or involve few integer variables, so the solver reaches (near-)optimal solutions quickly. Family F6, in contrast, introduces MOQ binary triggers and pack-size integer variables that create harder branch-and-bound trees; the 60-second limit may yield solutions with residual gaps.

\begin{table}[ht]
\centering
\small
\caption{Accuracy tolerances by problem structure.}
\label{tab:tolerances}
\begin{tabularx}{\textwidth}{@{}l>{\raggedright\arraybackslash}p{2.3cm}>{\raggedright\arraybackslash}p{2.5cm}>{\raggedright\arraybackslash}X@{}}
\toprule
\textbf{Families} & \textbf{Problem Type} & \textbf{Tolerance $\epsilon$} & \textbf{Rationale} \\
\midrule
F1--F5, F7--F8 & LP / easy MIP & $10^{-4}$ (strict) / $10^{-2}$ (practical) & LP relaxation tight; optimal is exact \\
F6 & Hard MIP (MOQ, pack size) & $10^{-2}$ (both tiers) & 60s time limit may yield near-optimal solutions; pack-size rounding creates inherent gaps \\
\bottomrule
\end{tabularx}
\end{table}

\subsection{Evaluation Metrics}

We report three primary metrics; precise mathematical definitions appear in Appendix~\ref{app:metrics_detail}:
\begin{itemize}[nosep]
    \item \textbf{Execution Rate (Exec\%)} = fraction of instances whose generated code runs without runtime error \emph{and} where the solver returns a feasible solution. Infeasibility, unboundedness, and runtime errors all count as execution failure.
    \item \textbf{Accuracy (Acc\%)} = fraction of instances satisfying both feasibility-status match and $|y_{\text{pred}} - y_{\text{ref}}|\,/\,|y_{\text{ref}}| < \epsilon$.
    \item \textbf{Silent Failure Rate (SF\%)} = $\text{Exec\%} - \text{Acc\%}$, the absolute gap between solver-feasible execution and correctness.
\end{itemize}
The silent failure rate is the primary metric of interest: it captures the absolute fraction of \emph{seemingly successful} runs that produce incorrect objective values due to semantic modeling errors.

\section{ReLoop Implementation Details}
\label{app:reloop_details}

This section provides complete implementation details for the two-layer verification framework described in Section~\ref{sec:verification}.
All prompts and parameters correspond to the source code in \texttt{reloop/}.

\begin{algorithm}[ht]
\caption{ReLoop: Behavioral Verification for LLM-Generated Optimization Code}
\label{alg:reloop}
\begin{algorithmic}[1]
\Require Problem description $x$, LLM $G$, iteration budget $N$
\Ensure Verified code $C$, objective $z^*$, solution $\mathbf{x}^*$, status, diagnostics $\mathcal{D}$
\Statex \hrulefill
\Statex \textbf{Phase 1: Structured Generation + L1 Verification}
\State $C \gets \textsc{StructuredGeneration}(G, x)$ \Comment{Eq.~\ref{eq:cot}}
\For{$i = 1, \ldots, N$} \Comment{L1 regeneration loop}
    \State $(\textit{status}, z^*, \mathbf{x}^*, \textit{diag}) \gets \textsc{L1-Verify}(C)$
    \If{$\textit{status} \neq \textsc{Fatal}$} \textbf{break}
    \EndIf
    \State $C \gets \textsc{Regenerate}(G, x, \textit{diag})$ \Comment{IIS/ray-guided}
\EndFor
\If{$\textit{status} = \textsc{Fatal}$}
    \Return $(C, \bot, \bot, \textsc{Failed}, \emptyset)$
\EndIf
\Statex \hrulefill
\Statex \textbf{Phase 2: L2 Behavioral Testing + Diagnosis-Guided Repair}
\For{$j = 1, \ldots, N$} \Comment{Diagnostic repair loop}
    \State $\mathcal{D} \gets \textsc{L2-CPT}(C, z^*, x) \;\cup\; \textsc{L2-OPT}(C, z^*, x)$
    \If{no $d \in \mathcal{D}$ has severity \textsc{Warning}}
        \Return $(C, z^*, \mathbf{x}^*, \textsc{Verified}, \mathcal{D})$ \Comment{Skip guard}
    \EndIf
    \State $C' \gets \textsc{TargetedRepair}(G, C, \mathcal{D})$
    \If{$\neg\,\textsc{SafetyCheck}(C')$} \Comment{Block data mutation}
        \State $C' \gets \textsc{GuidedRetry}(G, C, \mathcal{D})$ \Comment{Free retry, no budget consumed}
        \If{$\neg\,\textsc{SafetyCheck}(C')$} \textbf{break} \Comment{Halt repair on persistent unsafe output}
        \EndIf
    \EndIf
    \State $(\textit{status}', z'^*, \mathbf{x}'^*, \_) \gets \textsc{L1-Verify}(C')$
    \If{$\textsc{Regression}(z'^*, \textit{status}', z^*, \textit{status})$} \Comment{$|z'^* {-} z^*|/|z^*| > \tau_r$}
        \State \textbf{break} \Comment{Rollback: keep $C, z^*, \mathbf{x}^*$}
    \EndIf
    \State $C,\; z^*,\; \mathbf{x}^*,\; \textit{status} \gets C',\; z'^*,\; \mathbf{x}'^*,\; \textit{status}'$
\EndFor
\State \Return $(C, z^*, \mathbf{x}^*, \textit{status}, \mathcal{D})$
\end{algorithmic}
\end{algorithm}

\FloatBarrier
\subsection{Verification Layer Configuration}

Table~\ref{tab:reloop_params} summarizes the hyperparameters for each verification layer.

\begin{table}[ht]
\centering
\small
\caption{Complete parameter configuration for ReLoop verification layers.}
\label{tab:reloop_params}
\begin{tabularx}{\textwidth}{@{}lll>{\raggedright\arraybackslash}X@{}}
\toprule
\textbf{Layer} & \textbf{Parameter} & \textbf{Value} & \textbf{Description} \\
\midrule
\multirow{3}{*}{\textbf{L1} Execution}
  & \texttt{timeout} & 60\,s & Subprocess execution timeout \\
  & \texttt{max\_regenerations} & 3 & Regeneration attempts on \textsc{Fatal} \\
  & \texttt{duality\_gap\_threshold} & 0.01 & Primal--dual gap threshold (1\%, \textsc{Info} only) \\
\midrule
\multirow{3}{*}{\shortstack[l]{\textbf{L2} CPT\\(Constraint Presence)}}
  & \texttt{missing\_threshold} & 0.05 & $<5\%$ change $\to$ \textsc{Warning} \\
  & \texttt{uncertain\_threshold} & 0.30 & 5--30\% change $\to$ \textsc{Info} \\
  & \texttt{max\_candidates} & 10 & Max constraints to test per problem \\
\midrule
\multirow{3}{*}{\shortstack[l]{\textbf{L2} OPT\\(Objective Presence)}}
  & \texttt{missing\_threshold} & 0.05 & $<5\%$ change $\to$ \textsc{Warning} \\
  & \texttt{uncertain\_threshold} & 0.30 & 5--30\% change $\to$ \textsc{Info} \\
  & \texttt{max\_candidates} & 10 & Max objective terms to test per problem \\
\midrule
\multirow{2}{*}{\textbf{Pipeline}}
  & $N$ (repair budget) & 3 & Total repair iterations allowed \\
  & $\tau_r$ (regression guard) & 0.04 & Rollback if objective shifts $>$4\% \\
\bottomrule
\end{tabularx}
\end{table}

\subsection{Code Generation}
\label{app:generation}

ReLoop uses Chain-of-Thought (CoT) prompting with four reasoning stages in a single LLM call.
The generation pipeline uses an \emph{extraction-first} strategy with a self-contained fallback:

\begin{enumerate}[leftmargin=*, nosep]
    \item \textbf{Try extraction}: An LLM call extracts all numerical parameters from the problem description into a structured JSON dictionary. If successful, a CoT prompt instructs the model to reference this pre-loaded \texttt{data} variable (e.g., \texttt{data["capacity"]}) rather than embedding values.
    \item \textbf{Fallback}: If extraction fails (invalid JSON, empty result, or the generated code contains \texttt{json.loads}), the pipeline falls back to self-contained generation where the LLM embeds all data directly in the code.
\end{enumerate}

The extraction path enables L2 behavioral testing via data-dict perturbation. When extraction fails, L2 falls back to source-code AST perturbation (Section~\ref{app:perturbation_strategy}).

\paragraph{Chain-of-Thought Prompt (Self-Contained Fallback).}

The following prompt is used when data extraction fails. The data-reference variant shares the same four-step structure but adapts Steps~2, 3, and~4 for external data access (see note below).

\newpage
\begin{mdframed}[style=promptbox, frametitle={\small\sffamily Chain-of-Thought Generation Prompt}]
\begin{lstlisting}[style=prompt]
[System] You are an optimization expert who solves problems with
         step-by-step reasoning.

Solve this optimization problem using chain-of-thought reasoning.

## Problem
{problem_description}

---
## STEP 1: UNDERSTAND THE PROBLEM
First, analyze the problem:
- What is the objective? (minimize cost / maximize profit / etc.)
- What decisions need to be made?
- What constraints exist?
- What parameters are given?

## STEP 2: FORMULATE THE MATHEMATICAL MODEL
Write the formal model:
- Sets and indices
- Parameters (extract all numerical values from the problem
  description)
- Decision variables with domains
  **Variable Type**: For each variable, explicitly decide CONTINUOUS,
  INTEGER, or BINARY. Look for context where fractional values would
  be physically meaningless (e.g., number of trucks, workers to hire,
  items to select). State your choice and reasoning.
- Constraints in mathematical notation
- Objective function

## STEP 3: GENERATE GUROBI CODE
Write self-contained Python code using gurobipy.

**CRITICAL RULES:**
1. Define ALL data within your code (extract numbers from the problem
   description above)
2. Model variable must be named `m`
3. Set `m.Params.OutputFlag = 0`
4. Print exactly: `print(f"status: {m.Status}")` and
   `print(f"objective: {m.ObjVal}")`
5. Implement ALL constraints mentioned in the problem description
   (not just those in Step 2 -- re-read the problem to ensure
   nothing is missed)
6. Include ALL cost/revenue terms from the problem in the objective
   function

**Big-M Guidelines (if using indicator/logical constraints):**
- NEVER hardcode Big-M values like `M = 1e6`
- ALWAYS compute M dynamically from data parameters

**Edge Case Handling:**
- Check array length before iteration
- Avoid division by zero: `max(value, 1e-6)`

## STEP 4: VERIFY COMPLETENESS
Before finalizing, cross-check your code against the original problem:
- Does the objective include EVERY cost/revenue term mentioned in the
  problem?
- Is EVERY constraint from the problem implemented in the code?
- Are all numerical values correctly extracted from the problem
  description?
If anything is missing, fix the code before returning it.

---
Now solve the problem. Show your reasoning for Steps 1-2, then
provide the final code in a ```python block.
\end{lstlisting}
\end{mdframed}

\paragraph{Data-Reference Variant.}
When extraction succeeds, the prompt differs in three steps:
\begin{itemize}[leftmargin=*, nosep]
    \item \textbf{Step~2}: ``\textit{Parameters (reference the data keys listed below)}'' replaces ``\textit{Parameters (extract all numerical values from the problem description)}''.
    \item \textbf{Step~3}: ``\textit{Write Python code using gurobipy}'' (no ``self-contained''). Rule~1 becomes ``\textit{The \texttt{data} variable is PRE-LOADED with the problem data. Do NOT define or redefine \texttt{data}. Just use \texttt{data["key"]} directly.}'' An additional Rule~7 is added: ``\textit{Do NOT use \texttt{import json} or \texttt{json.loads()}. Data is already a Python dict.}''
    \item A data schema section (``\texttt{\#\# Available Data Keys}'') is inserted between Steps~3 and~4, listing all keys with types and dimensions (but not values).
    \item \textbf{Step~4}: ``\textit{Are data keys accessed correctly?}'' replaces ``\textit{Are all numerical values correctly extracted from the problem description?}''
\end{itemize}

\subsection{L1: Execution Verification}

L1 is a blocking layer that verifies code executability. It performs four sequential checks:

\begin{enumerate}[leftmargin=*, nosep]
    \item \textbf{Syntax check}: Compile-time validation via \texttt{compile()}
    \item \textbf{Execution}: Run code in a sandboxed subprocess with 60\,s timeout
    \item \textbf{Solver status}: Check for \textsc{Infeasible} (with IIS diagnostics), \textsc{Unbounded} (with unbounded ray variables), or \textsc{Timeout}
    \item \textbf{Duality check}: If \textsc{Optimal}, compare primal--dual gap. Gap $>1\%$ emits \textsc{Info} (does \emph{not} trigger repair)
\end{enumerate}

A failure in any of the first three checks emits a \textsc{Fatal} diagnostic and triggers the regeneration loop (up to 3 attempts).

\paragraph{L1 Regeneration Prompt.}

When L1 detects a \textsc{Fatal} error, the pipeline regenerates code using the error message as feedback. The data instructions section is \emph{conditional}: if the code uses an external data dict, it includes the data schema and access rules; if the code is self-contained, it states ``\textit{Code is self-contained --- all data is defined within the code itself.}''

\begin{mdframed}[style=promptbox, frametitle={\small\sffamily L1 Regeneration Prompt}]
\begin{lstlisting}[style=prompt]
[System] You fix broken optimization code. Ensure the new code is
         syntactically correct and handles all edge cases.

The previous code failed to execute. Generate a new, correct version.

## Problem
{problem_description}

## Previous Code (FAILED)
```python
{failed_code}
```

## Error
{error_message}

{data_instructions}

## Instructions
1. Analyze why the previous code failed
2. Generate completely new code that avoids the error
3. Handle edge cases (empty arrays, division by zero)

Return ONLY the corrected Python code in a ```python block.
\end{lstlisting}
\end{mdframed}

Where \texttt{\{data\_instructions\}} expands to either:
\begin{itemize}[nosep]
    \item \textbf{Data-dict mode}: ``\texttt{\#\# Data Structure} / The \texttt{data} variable is PRE-DEFINED with these keys: \{schema\} / \textbf{CRITICAL}: Do NOT create \texttt{data = \{...\}}. Just use \texttt{data["key"]} directly.''
    \item \textbf{Self-contained mode}: ``\texttt{\#\# Note} / Code is self-contained --- all data is defined within the code itself.''
\end{itemize}

\subsection{L2: Constraint Presence Testing (CPT)}

L2 CPT tests whether constraints expected from the problem description are actually enforced in the generated code, using solver-based perturbation.

\paragraph{Constraint Extraction Prompt.}

An LLM extracts candidate constraints from the problem description. The prompt provides the list of available data parameter keys (but not values) to help the LLM identify relevant parameters.

\begin{mdframed}[style=promptbox, frametitle={\small\sffamily L2 CPT Constraint Extraction Prompt}]
\begin{lstlisting}[style=prompt]
Analyze this optimization problem and extract the KEY CONSTRAINTS
that should be present in the model.

## Problem Description
{problem_description}

## Available Data Parameters
{list(data.keys())}

## Task
Identify constraints that are REQUIRED by the problem. Focus on:
1. Capacity constraints (resource limits, maximum values)
2. Demand constraints (minimum requirements, must-satisfy conditions)
3. Balance constraints (flow balance, inventory balance)

## Output Format
Return ONLY a JSON array with this exact format:
```json
[
  {"description": "minimum protein requirement",
   "type": "demand", "parameters": ["min_protein"]},
  {"description": "capacity limit on production",
   "type": "capacity", "parameters": ["capacity"]}
]
```

Return ONLY the JSON array, no explanation.
\end{lstlisting}
\end{mdframed}

\paragraph{Testing Procedure.}
\label{app:perturbation_strategy}

For each candidate constraint (up to 10), the pipeline applies extreme perturbation to the associated parameter and measures objective change. A \emph{dual-strategy} perturbation approach is used:

\begin{enumerate}[leftmargin=*, nosep]
    \item \textbf{Strategy~1 (data-dict)}: Perturb the parameter in the external data dictionary and re-execute the original code. Used when the code reads from the \texttt{data} variable.
    \item \textbf{Strategy~2 (source-code fallback)}: If the code embeds data directly in source (detected automatically), or if Strategy~1 produced $<1\%$ objective change in hybrid mode, the pipeline falls back to AST-based source-code perturbation: it locates the parameter's assignment in the code via fuzzy name matching, modifies the literal value, and re-executes.
\end{enumerate}

The perturbation factor depends on constraint type:
\begin{itemize}[nosep]
    \item Capacity constraints: $\times 0.001$ (near-zero)
    \item Demand constraints: $\times 100$ (extreme increase)
    \item Other constraints: $\times 0.01$ (1\% of original)
\end{itemize}

Result classification:
\begin{enumerate}[leftmargin=*, nosep]
    \item If perturbation causes \textsc{Infeasible}: constraint is present $\to$ \textsc{Pass}
    \item Compute change ratio: $r = |z_{\text{new}} - z^*| \,/\, |z^*|$ \quad (absolute change used when $|z^*| < \varepsilon$)
    \item Classify (consistent with \S\ref{sec:l2}):
    \begin{itemize}[nosep]
        \item $r < \tau_\ell = 5\%$: Constraint likely missing $\to$ \textsc{Warning} (triggers repair)
        \item $\tau_\ell \le r \le \tau_h = 30\%$: Uncertain $\to$ \textsc{Info} (no repair)
        \item $r > \tau_h$ or infeasibility: Constraint present $\to$ \textsc{Pass}
    \end{itemize}
\end{enumerate}

\subsection{L2: Objective Presence Testing (OPT)}

L2 OPT tests whether expected cost and revenue terms are present in the generated objective function, using the same perturbation methodology as CPT.

\newpage
\paragraph{Objective Term Extraction Prompt.}

\begin{mdframed}[style=promptbox, frametitle={\small\sffamily L2 OPT Extraction Prompt}]
\begin{lstlisting}[style=prompt]
Analyze this optimization problem and extract the KEY OBJECTIVE
FUNCTION TERMS (cost and revenue components) that should be present
in the model's objective function.

## Problem Description
{problem_description}

## Available Data Parameters
{list(data.keys())}

## Task
Identify cost and revenue terms that MUST appear in the objective
function. Focus on:
1. **Cost terms**: purchasing/procurement cost, holding/storage cost,
   transportation cost, shortage/backorder cost, setup/fixed cost,
   penalty cost
2. **Revenue terms**: sales revenue, demand revenue, return/salvage
   value

For each term, identify which data parameter(s) provide its
coefficient.

## Output Format
Return ONLY a JSON array with this exact format:
```json
[
  {"description": "unit purchasing cost",
   "role": "cost", "parameters": ["unit_cost"]},
  {"description": "sales revenue per unit",
   "role": "revenue", "parameters": ["selling_price"]}
]
```

Return ONLY the JSON array, no explanation.
\end{lstlisting}
\end{mdframed}

\paragraph{Testing Procedure.}
For each candidate objective term (up to 10), the same dual-strategy perturbation is applied. The perturbation factor depends on the term's role:
\begin{itemize}[nosep]
    \item Cost terms: $\times 0.001$ (near-zero)
    \item Revenue terms: $\times 100$ (extreme increase)
    \item Other terms: $\times 0.01$ (1\% of original)
\end{itemize}

Classification follows the same thresholds as CPT ($<5\%$ \textsc{Warning}, 5--30\% \textsc{Info}, $\ge 30\%$ \textsc{Pass}), except that infeasibility is not expected from objective perturbation and is treated as a skipped test (no result emitted).

\subsection{Repair Pipeline}

The repair loop processes unified \texttt{Diagnostic} objects produced by L1 and L2. Diagnostics are split into \emph{actionable} issues (\texttt{triggers\_repair=True}, severity \textsc{Warning} or above) and \emph{reference-only} items (\textsc{Info}, likely normal). The repair loop runs up to $N=3$ iterations and terminates early if:

\begin{itemize}[nosep]
    \item Status is \textsc{Verified} with no actionable diagnostics (skip repair entirely)
    \item Repair produces identical code (no change)
    \item Repair does not change status or objective (plateau)
    \item Regression detected: objective shifts $>4\%$, status degrades, or solution crashes
\end{itemize}

\paragraph{Repair Prompt.}

The repair prompt is assembled dynamically by \texttt{build\_repair\_prompt()} from the list of \texttt{Diagnostic} objects. The data section and safety rules adapt based on whether the code uses an external data dict or is self-contained.

\newpage
\begin{mdframed}[style=promptbox, frametitle={\small\sffamily Repair Prompt (build\_repair\_prompt)}]
\begin{lstlisting}[style=prompt]
[System] You are an optimization code repair expert.

CRITICAL RULES:
1. ONLY fix the actionable issues listed in the ISSUES DETECTED
   section
2. Items in REFERENCE ONLY are for context -- DO NOT modify code
   based on them
3. Be conservative -- only make changes that are clearly necessary
4. Preserve all working code -- only change what is broken
5. Do NOT change hardcoded data values unless the diagnostic
   evidence specifically requires it

Fix this optimization code based on the behavioral verification
report.

## Problem
{problem_description}

{data_section}

## Current Code
```python
{code}
```

## Current objective value: {current_obj}

---
## ISSUES DETECTED ({N} actionable)

=== Issue 1 [{layer}] [{severity}] ===
Type: {issue_type}
Target: {target_name}
Evidence: {evidence}

=== Issue 2 ... ===

---
## REFERENCE ONLY (DO NOT FIX)

+-----------------------------------------------------------------+
| Below items are NORMAL in 80%+ of cases.  DO NOT CHANGE.       |
+-----------------------------------------------------------------+

1. [{layer}] {issue_type} -- {target_name}
   {evidence}
   Action: DO NOT FIX (unless 100% certain this is an error)

---
## REPAIR INSTRUCTIONS

1. Read each Issue carefully, especially the Evidence field
2. Identify the root cause in your code for each actionable issue
3. Fix ALL actionable issues above
4. DO NOT fix items in the REFERENCE section -- they are likely
   normal
5. Preserve all working code -- only change what is broken

{safety_rules}

Return the COMPLETE fixed code in a ```python block.
\end{lstlisting}
\end{mdframed}

Where \texttt{\{data\_section\}} and \texttt{\{safety\_rules\}} are conditional:

\begin{itemize}[nosep]
    \item \textbf{Data-dict mode}: The data section shows the schema (keys, types, dimensions but not values). Safety rules prohibit redefining \texttt{data}, using \texttt{json.loads()}, and mutating data contents.
    \item \textbf{Self-contained mode}: The data section states ``Code is self-contained.'' Safety rules prohibit removing or changing hardcoded data values unless diagnostic evidence requires it.
\end{itemize}

\paragraph{Safety Guardrails.}

Before accepting repaired code, a safety validator (\texttt{repair\_safety.py}) checks for dangerous operations using both regex pattern matching and AST analysis:

\begin{enumerate}[leftmargin=*, nosep]
    \item \textbf{Data reassignment}: Redefining \texttt{data} with a dict literal (\texttt{data = \{...\}}) is blocked. Exception: \texttt{data = json.loads(...)} is allowed as it re-parses existing data rather than fabricating values.
    \item \textbf{Data mutation}: Modifying data contents (\texttt{data["key"] = value}) is blocked.
    \item \textbf{Dangerous imports}: \texttt{os} and \texttt{subprocess} modules are blocked.
\end{enumerate}

If violations are detected, a guided re-repair is attempted once (prepending violation details to the prompt). This safety retry does \emph{not} consume the repair budget. If the retry also fails validation, the original (pre-repair) code is kept.

\section{Evaluation Details}
\label{app:evaluation}

\subsection{Correctness Criteria}

An instance is \textbf{correct} if:
\begin{enumerate}[leftmargin=*, nosep]
    \item \textbf{Status match:} Predicted feasibility equals ground truth
    \item \textbf{Objective match:} For feasible instances, relative error $< \epsilon$
\end{enumerate}

\begin{table}[ht]
\centering
\small
\caption{Tolerance thresholds by benchmark and family.}
\begin{tabular}{@{}llll@{}}
\toprule
\textbf{Benchmark} & \textbf{Families} & \textbf{Problem Type} & \textbf{Tolerance $\epsilon$} \\
\midrule
RetailOpt-190 & F1--F5, F7--F8 & LP & $10^{-4}$ (strict) / $10^{-2}$ (practical) \\
RetailOpt-190 & F6 & MIP & $10^{-2}$ (both tiers) \\
MAMO-ComplexLP & -- & LP & $10^{-6}$ \\
IndustryOR & -- & Mixed & $10^{-6}$ \\
\bottomrule
\end{tabular}
\end{table}

RetailOpt-190 reports accuracy at two tolerance tiers to capture different aspects of formulation quality: $\epsilon = 10^{-4}$ measures strict formulation correctness (exact match), while $\epsilon = 10^{-2}$ captures practically correct solutions with minor numerical discrepancies. For MIP family~F6, both tiers use $\epsilon = 10^{-2}$ since integer variables under the 60-second time limit may yield near-optimal rather than provably optimal solutions. Cross-benchmark experiments adopt $\epsilon = 10^{-6}$ following the SIRL evaluation protocol~\citep{chen2025solverinformedrl}.

\subsection{Metrics}
\label{app:metrics_detail}

All metrics are computed over the full dataset of $N$ instances.

\paragraph{Primary Metrics.}
\begin{align}
\text{Exec\%} &= \frac{|\{i : C_i \text{ executes} \wedge 
                 \text{solver returns feasible } z_i\}|}{N} \times 100 \\
\text{Acc\%}(\epsilon)  &= \frac{|\{i : |z_i - z_i^{\text{gt}}| / 
                 |z_i^{\text{gt}}| < \epsilon \}|}{N} \times 100
\end{align}

We report Acc\% at two tolerance levels for RetailOpt-190: Acc\%($\epsilon\!=\!10^{-4}$) and Acc\%($\epsilon\!=\!10^{-2}$). For cross-benchmark experiments, we report Acc\%($\epsilon\!=\!10^{-6}$) only. All results use pass@1 (single attempt, greedy decoding).

\paragraph{Derived Metrics (Appendix Tables).}
For analyses in the appendix, we also compute:
\begin{align}
\text{SF\%}   &= \text{Exec\%} - \text{Acc\%} \quad \text{(silent failure rate)}
\end{align}

An instance is considered a \emph{silent failure} if it executes and the solver returns a feasible solution, but the objective does not match ground truth within tolerance. SF\% quantifies the feasibility--correctness gap discussed in \S\ref{sec:main_results}.

\section{Experimental Setup}
\label{app:exp_details}

\subsection{Model Configuration}

\begin{table}[ht]
\centering
\small
\caption{Model configurations used in experiments. All models use greedy decoding for reproducibility.}
\label{tab:llm_config}
\begin{tabular}{@{}llllll@{}}
\toprule
\textbf{Type} & \textbf{Model} & \textbf{Provider} & \textbf{Temp.} & \textbf{Max Tokens} & \textbf{Notes} \\
\midrule
Foundation & Claude~Opus~4.6 & Anthropic API & 0.0 & 8192 & \\
Foundation & DeepSeek-V3.2 & DeepSeek API & 0.0 & 8192 & \\
Foundation & Qwen3-32B & Local (vLLM~\citep{kwon2023vllm}) & 0.0 & 8192 & BF16 \\
Offline SFT & OptMATH-Qwen2.5-32B & Local (vLLM) & 0.0 & 8192 & BF16 \\
Online RL & SIRL-Qwen2.5-32B & Local (vLLM) & 0.0 & 8192 & BF16 \\
\bottomrule
\end{tabular}
\end{table}

\paragraph{Evaluation Configurations.}
All five models are evaluated under three configurations: 
\emph{Base/Native} (direct generation for foundation models, own 
fine-tuned format for SFT/RL), \emph{CoT} (structured chain-of-thought 
with Understand $\to$ Formalize $\to$ Synthesize $\to$ Verify stages), and 
\emph{ReLoop} (CoT + two-layer behavioral verification with 
diagnosis-guided repair, $N\!=\!3$).

\paragraph{L2 Verification LLM.}
For L2 behavioral testing (CPT and OPT), the constraint and objective extraction uses the same LLM that generates the code, with temperature~$= 0$; cross-model extraction is left for future work.

\subsection{Solver Configuration}

Ground truth computed with Gurobi 11.0:
\begin{itemize}[leftmargin=*, nosep]
    \item \texttt{TimeLimit}: 60 seconds
    \item \texttt{MIPGap}: 0.01 (1\% optimality gap for MIP)
    \item \texttt{Threads}: 1 (single-threaded for reproducibility)
    \item \texttt{Seed}: 0 (fixed random seed)
    \item \texttt{OutputFlag}: 0 (suppress solver logs)
\end{itemize}

\subsection{Benchmark Details}

\begin{table}[ht]
\centering
\small
\caption{Benchmark statistics.}
\label{tab:benchmark_stats}
\begin{tabular}{@{}lcccl@{}}
\toprule
\textbf{Benchmark} & \textbf{Instances} & \textbf{Avg Tokens} & \textbf{Tolerance} & \textbf{Format} \\
\midrule
RetailOpt-190 & 190 & $\sim$2{,}900 & $10^{-4}$ / $10^{-2}$ & Data-embedded \\
MAMO-ComplexLP & 203 & $\sim$459 & $10^{-6}$ & Data-embedded \\
IndustryOR & 100 & $\sim$267 & $10^{-6}$ & Data-embedded \\
\bottomrule
\end{tabular}
\end{table}

All benchmarks use data-embedded format (full data in prompt) for evaluation, ensuring comparability with prior work. RetailOpt-190 also provides a schema-based format used internally by the ReLoop verification pipeline for parameter perturbation.

\paragraph{Baselines.}
All Base, CoT, and ReLoop values in Tables~\ref{tab:cross} and~\ref{tab:ablation} come from our own runs under the protocol of this appendix, including the Base configurations of DeepSeek-V3.2, Qwen3-32B, OptMATH-32B, and SIRL-32B. For OptMATH-32B and SIRL-32B, our reproduced Base accuracies differ from the values reported by \citet{chen2025solverinformedrl}; we report our reproductions so that every configuration is evaluated under the same protocol. For context, \citet{chen2025solverinformedrl} (Table~1) report the following results, which we did not rerun: GPT-4 (49.3\% / 33.0\%), DeepSeek-R1 (67.9\% / 45.0\%), and OpenAI-o3 (51.2\% / 44.0\%) on MAMO-ComplexLP / IndustryOR.

\subsection{Reproducibility}

All LLM calls use greedy decoding (temperature~$= 0$), ensuring 
deterministic outputs. Solver random seed is fixed at 0. All results 
are single-run (pass@1); we verified reproducibility on a random subset of 20 
instances, observing identical outputs across repeated executions.

\section{Per-Family Analysis on RetailOpt-190}
\label{app:family_analysis}

Table~\ref{tab:family_results} presents per-family accuracy for all five models. This breakdown reveals which constraint interaction patterns are most challenging and where ReLoop provides the largest gains.

\begin{table}[ht]
\centering
\small
\caption{Per-family results on RetailOpt-190 (Acc\%, $\epsilon\!=\!10^{-4}$). Base = direct generation for foundation models, native format for SFT/RL models. +ReLoop = CoT + L1--L2 for all models.}
\label{tab:family_results}
\vspace{0.5em}
\setlength{\tabcolsep}{1.6pt}
\begin{tabular}{@{}l c cc cc cc cc cc@{}}
\toprule
& & \multicolumn{2}{c}{\textbf{Claude~Opus~4.6}} & \multicolumn{2}{c}{\textbf{DeepSeek~V3.2}} & \multicolumn{2}{c}{\textbf{Qwen3-32B}} & \multicolumn{2}{c}{\textbf{OptMATH-32B}} & \multicolumn{2}{c}{\textbf{SIRL-32B}} \\
\cmidrule(lr){3-4} \cmidrule(lr){5-6} \cmidrule(lr){7-8} \cmidrule(lr){9-10} \cmidrule(lr){11-12}
\textbf{Family} & \textbf{\#} & Base & +ReLoop & Base & +ReLoop & Base & +ReLoop & Base & +ReLoop & Base & +ReLoop \\
\midrule
\textbf{F1 Core Ops}        & \textbf{20} & 55.0 & \textbf{95.0} & 5.0 & \textbf{5.0} & 0.0 & 0.0 & 0.0 & 0.0 & 0.0 & 0.0 \\
\textbf{F2 Assort \& Sub}   & \textbf{30} & 50.0 & \textbf{53.3} & 0.0 & \textbf{3.3} & 0.0 & 0.0 & 0.0 & 0.0 & 0.0 & 0.0 \\
\textbf{F3 Resource}         & \textbf{20} & 0.0 & 0.0 & 0.0 & 0.0 & 0.0 & 0.0 & 0.0 & 0.0 & 0.0 & 0.0 \\
\textbf{F4 Demand Dyn}       & \textbf{30} & 3.3 & \textbf{20.0} & 0.0 & \textbf{10.0} & 0.0 & 0.0 & 0.0 & 0.0 & 0.0 & 0.0 \\
\textbf{F5 Feasibility}      & \textbf{20} & 0.0 & 0.0 & 0.0 & 0.0 & 0.0 & 0.0 & 0.0 & 0.0 & 0.0 & 0.0 \\
\textbf{F6 Discrete Log}     & \textbf{20} & 0.0 & 0.0 & 0.0 & \textbf{20.0} & 0.0 & 0.0 & 0.0 & 0.0 & 0.0 & 0.0 \\
\textbf{F7 Network \& ME}    & \textbf{30} & 20.0 & \textbf{26.7} & 0.0 & 0.0 & 0.0 & 0.0 & 0.0 & 0.0 & 0.0 & 0.0 \\
\textbf{F8 Omni-channel}     & \textbf{20} & \textbf{50.0} & \textbf{50.0} & 0.0 & \textbf{10.0} & 0.0 & 0.0 & 0.0 & 0.0 & 0.0 & 0.0 \\
\midrule
\textbf{Total}     & \textbf{190} & \textbf{22.6} & \textbf{31.1} & \textbf{0.5} & \textbf{5.8} & \textbf{0.0} & \textbf{0.0} & \textbf{0.0} & \textbf{0.0} & \textbf{0.0} & \textbf{0.0} \\
\bottomrule
\end{tabular}
\end{table}

The breakdown shows four patterns.
F1 (Core Operations) shows the largest absolute gain for Claude (55.0\%$\to$95.0\%), which suggests that ReLoop is most effective on basic perishable inventory dynamics.
F3 (Resource) and F5 (Feasibility Stress) remain at 0\% across all models, suggesting that multi-resource coupling and deliberately tight feasibility boundaries exceed current LLM modeling capabilities regardless of verification.
In F6 (Discrete Logistics), Claude gains nothing while DeepSeek rises from 0\% to 20.0\%, which suggests that L1 recovery from crashes caused by integer-variable errors accounts for DeepSeek's gain on these MIP problems.
The 32B models (Qwen3, OptMATH, SIRL) score 0\% in every family, which is consistent with these models lacking the capacity for compositional retail optimization at this level of complexity.

\section{Injected-Defect Characterization of Behavioral Verification}
\label{app:defect_study}

To characterize L2 against construction-known labels, we generated 240 single-component knockout mutants of the RetailOpt-190 reference formulations across all 38 archetypes (one constraint family or objective term removed per mutant via disable switches in the reference solver), plus 76 structural-tier mutants (aging linkage, initial-condition) reported separately. Candidate constraints and objective terms were generated programmatically from the instance schema (\emph{oracle extraction}), decoupling extraction error from detection error; solver settings match the evaluation protocol. None of the 240 omissions causes infeasibility, because deleting a constraint only relaxes the feasible region and deleting an objective term does not affect feasibility. Omission defects are therefore invisible to L1 by construction and form exactly the silent-failure class that L2 targets, whereas commission errors (over-constraining, missing slack, broken syntax) are handled by L1.

\textbf{Detection.} L2 detects 238/240 (99.2\%): 83/83 removed constraints (CPT) and 155/157 removed objective terms (OPT), including 74/74 components inactive at nominal data and detectable only because extreme perturbation forces them to bind. The two misses share one mechanism: the purchasing coefficient also enters the per-period budget constraint, so perturbing it moves the objective through that constraint even with the objective term absent, which is the parameter-sharing case of Table~\ref{tab:scope}. The intact-code false-positive rate is 26.7\% of candidates (64/240; CPT 4/83, OPT 60/157), concentrated on objective terms and arising from a single mechanism: components genuinely present but economically inactive even under perturbation (e.g., cost terms of products the optimum never purchases). Its practical cost is limited by conservative repair (re-adding a present constraint leaves the feasible region unchanged), \textsc{Info}-level non-repair, and the regression guard.

\textbf{Multiplier robustness and taxonomy independence.} Detection recall is 238/240 at each shrink multiplier, $10^{-4}$, $10^{-3}$, and $10^{-2}$, which span three orders of magnitude, and intact-code false positives are 63--64/240 (Table~\ref{tab:grid}). Thirty-eight knockouts have zero-valued optima (lost-sales term removal) and are detected via the absolute-change fallback of Assumption~\ref{prop:perturb}. The inflate axis has no test population on this benchmark (every removable component is capacity- or cost-typed), so robustness is scoped to shrink-direction perturbations. A taxonomy-free control (bidirectional extreme perturbation per parameter, no type labels, max response) attains 238/240 with fewer intact-code false positives (54/240), so detection does not depend on the capacity/demand/cost routing, whose only role is to halve the number of solver calls.

\begin{table}[h]
\centering
\small
\caption{Injected-defect detection across shrink multipliers (240 knockouts; 38 intact baselines).}
\label{tab:grid}
\vspace{0.4em}
\begin{tabular}{@{}lccc c@{}}
\toprule
 & \multicolumn{3}{c}{\textbf{Typed routing}} & \textbf{Taxonomy-free} \\
\cmidrule(lr){2-4}
 & $\times 10^{-4}$ & $\times 10^{-3}$ & $\times 10^{-2}$ & (bidirectional) \\
\midrule
Recall & 238/240 & 238/240 & 238/240 & 238/240 \\
Intact-code FP & 63/240 & 64/240 & 64/240 & 54/240 \\
\bottomrule
\end{tabular}
\end{table}

\section{Failure-Attribution and Label-Validity Audit}
\label{app:audit}

We audited a stratified sample of 55 residual failures (RetailOpt 40 by family; MAMO 15 by $|$GT$|$ tertile) in two stages. First, an LLM pre-annotated the failures in two independent blind runs (raw agreement 76\%, $\kappa = 0.57$). Second, an expert verified the labels, and a second author independently re-verified every non-modeling-error label and every disagreement through independent Gurobi re-solves and adversarial tests against the official reference generator. Every label carries a solver-checkable minimal-flip rationale with recomputed objectives.

\textbf{RetailOpt-190.} Of the 40 sampled failures, 39 are true modeling errors and 1 is a numerical or tolerance case. None is a defensible alternative reading: we tested each hypothesized reading by forcing the alternative structure in the reference generator, and the resulting optimum matched neither the ground truth nor the model output. For example, toggling substitution leaves the ground truth at 528{,}518, while the model returns 488{,}753, 7.5\% below the true optimum, which indicates a relaxed or missing constraint. None is an equivalent formulation penalized by the metric. Of the 43 true modeling errors in the audited sample (both benchmarks), 42 are structural, namely wrong inter-mechanism coupling (22) or missing structural linkage (20), and one is a data-extraction slip. The reference generator reproduces all 190 published ground truths to $10^{-6}$ and matches Appendix~\ref{app:reference_milp} line by line.

\textbf{MAMO-ComplexLP.} We tested and rejected the hypothesis that integer-valued ground truths on diet problems with divisible-looking quantities are mislabeled. MAMO applies a consistent convention: countable goods carry integer ground truths (197/203 of ground truths are integer-valued), and genuinely continuous quantities carry fractional ones (all 6 fractional ground truths use continuous units). Independent LP+IP re-solves of all 26 integer-labeled diet instances confirm that each integer value is the intended optimum over discrete items. This convention agrees with the variable-type reasoning of Stage~2, and in all 11 L2-corrected MAMO instances the diagnosis-guided repair restored the integrality dropped by continuous relaxation (variable types changed; constraint sets unchanged); none of these instances involves a modified label. We also disclose that one instance was corrupted by our own preprocessing (ground truth 162 recorded as a $-9999$ sentinel; restored; no effect on scores) and that two isolated data errata are documented, with an effect on the scores of at most two instances.

\section{Benchmark Composition Audit}
\label{app:composition}

Under the standard definition of revenue management, the allocation or pricing of fixed perishable capacity (booking control, fare classes, markdown pricing), neither MAMO-ComplexLP nor IndustryOR contains revenue-management instances. MAMO's own category metadata labels every instance \texttt{complex\_lp}, and most problems in both sets are production, blending, transportation or network-flow, scheduling, and assignment problems. A minority carry revenue- or profit-maximization objectives (11/203 in MAMO; 33/100 in IndustryOR), which does not place a problem in the revenue-management class, and the single keyword hit (``contract bookings'') was judged by a human reviewer to be a false positive. The MAMO instances corrected by L2 are cost-minimization diet problems with no retail or revenue-management character; Appendix~\ref{app:defect_study} measures the detector across the mechanism families of RetailOpt-190.

\FloatBarrier
\newpage
\section*{NeurIPS Paper Checklist}

\begin{enumerate}

\item {\bf Claims}
    \item[] Question: Do the main claims made in the abstract and introduction accurately reflect the paper's contributions and scope?
    \item[] Answer: \answerYes{}
    \item[] Justification: The abstract and Section~1 state four contributions: structured generation as the largest source of accuracy gains for foundation models on compositional problems, behavioral verification as the largest single contributor on localized defects, the ReLoop pipeline with regression-guarded repair, and the RetailOpt-190 benchmark. Sections~\ref{sec:main_results}--\ref{sec:ablation} and Appendices~\ref{app:family_analysis}--\ref{app:composition} report the supporting results, and the Limitations paragraph in Section~\ref{sec:conclusion} states the scope.

\item {\bf Limitations}
    \item[] Question: Does the paper discuss the limitations of the work performed by the authors?
    \item[] Answer: \answerYes{}
    \item[] Justification: The Limitations paragraph in Section~\ref{sec:conclusion} states four limitations: the supported formulation class and the shrink-only multiplier evidence, the conflict between CoT and the output format of SFT models, possible error correlation from shared LLM extraction, and the detection boundaries listed in Table~\ref{tab:scope}.

\item {\bf Theory assumptions and proofs}
    \item[] Question: For each theoretical result, does the paper provide the full set of assumptions and a complete (and correct) proof?
    \item[] Answer: \answerNA{}
    \item[] Justification: The paper states no theorems. Assumption~\ref{prop:perturb} is a design assumption with explicit conditions and edge-case guards, and it is evaluated empirically in Section~\ref{sec:ablation} and Appendix~\ref{app:defect_study}.

    \item {\bf Experimental result reproducibility}
    \item[] Question: Does the paper fully disclose all the information needed to reproduce the main experimental results of the paper to the extent that it affects the main claims and/or conclusions of the paper (regardless of whether the code and data are provided or not)?
    \item[] Answer: \answerYes{}
    \item[] Justification: Section~\ref{sec:method} and Algorithm~\ref{alg:reloop} specify the pipeline; Appendix~\ref{app:reloop_details} gives all prompts and verification parameters; Appendices~\ref{app:reference_milp}--\ref{app:data_format} specify the benchmark; and Appendices~\ref{app:evaluation} and~\ref{app:exp_details} give the metrics, tolerances, model versions, decoding parameters, and solver settings.

\item {\bf Open access to data and code}
    \item[] Question: Does the paper provide open access to the data and code, with sufficient instructions to faithfully reproduce the main experimental results, as described in supplemental material?
    \item[] Answer: \answerYes{}
    \item[] Justification: The ReLoop code and the RetailOpt-190 benchmark are public at \url{https://github.com/junbolian/ReLoop} under the license included in the repository; Appendices~\ref{app:data_format} and~\ref{app:reloop_details} document the data format, prompts, and configuration used to produce the results.

\item {\bf Experimental setting/details}
    \item[] Question: Does the paper specify all the training and test details (e.g., data splits, hyperparameters, how they were chosen, type of optimizer) necessary to understand the results?
    \item[] Answer: \answerYes{}
    \item[] Justification: No models are trained. All evaluations use greedy decoding (temperature~$=0$, max\_tokens~$=8192$) and repair budget $N=3$; the L2 perturbation factors (capacity $\times 0.001$, demand $\times 100$, cost $\times 0.001$, revenue $\times 100$) are in Appendix~\ref{app:reloop_details}, and the tolerances ($\epsilon \in \{10^{-4}, 10^{-2}, 10^{-6}\}$), benchmark statistics, and Gurobi 11.0 settings are in Appendices~\ref{app:evaluation} and~\ref{app:exp_details}.

\item {\bf Experiment statistical significance}
    \item[] Question: Does the paper report error bars suitably and correctly defined or other appropriate information about the statistical significance of the experiments?
    \item[] Answer: \answerYes{}
    \item[] Justification: Section~\ref{sec:main_results} reports a paired exact sign test for the CoT gain ($p \approx 0.002$) and an archetype-level cluster bootstrap 95\% confidence interval ($[1.6, 16.8]$pp), and Section~\ref{sec:cross_benchmark} reports a sign test for the L2 gain on MAMO ($p \approx 0.02$). All runs are single-run pass@1 with greedy decoding; repeated runs on 20 instances produced identical outputs (Appendix~\ref{app:exp_details}).

\item {\bf Experiments compute resources}
    \item[] Question: For each experiment, does the paper provide sufficient information on the computer resources (type of compute workers, memory, time of execution) needed to reproduce the experiments?
    \item[] Answer: \answerYes{}
    \item[] Justification: Claude Opus~4.6 and DeepSeek-V3.2 are accessed through commercial APIs, and ReLoop adds about $3\times$ the base token cost (Section~\ref{sec:exp_setup}). The 32B models run locally through vLLM in BF16 (Appendix~\ref{app:exp_details}); per-instance inference completes in seconds on a single high-memory GPU, and single-threaded Gurobi 11.0 solves complete in seconds.
    
\item {\bf Code of ethics}
    \item[] Question: Does the research conducted in the paper conform, in every respect, with the NeurIPS Code of Ethics \url{https://neurips.cc/public/EthicsGuidelines}?
    \item[] Answer: \answerYes{}
    \item[] Justification: The research involves no human subjects and no personal data, and the released artifacts are synthetic optimization scenarios and a verification framework.

\item {\bf Broader impacts}
    \item[] Question: Does the paper discuss both potential positive societal impacts and negative societal impacts of the work performed?
    \item[] Answer: \answerYes{}
    \item[] Justification: \emph{Positive}: more reliable LLM-based optimization reduces the risk of confident but wrong recommendations in supply chain, inventory, and logistics decisions. \emph{Negative}: users may place more trust in verified pipelines than their detection scope supports; Table~\ref{tab:scope} states which errors L2 cannot detect, and the 90-point feasibility--correctness gap in Section~\ref{sec:main_results} is a caution against deploying unverified LLM-generated optimization code.
    
\item {\bf Safeguards}
    \item[] Question: Does the paper describe safeguards that have been put in place for responsible release of data or models that have a high risk for misuse (e.g., pre-trained language models, image generators, or scraped datasets)?
    \item[] Answer: \answerNA{}
    \item[] Justification: RetailOpt-190 contains synthetic optimization scenarios with no privacy, security, or dual-use concerns, and ReLoop is a verification framework rather than a generative model; no high-risk artifacts are released.

\item {\bf Licenses for existing assets}
    \item[] Question: Are the creators or original owners of assets (e.g., code, data, models), used in the paper, properly credited and are the license and terms of use explicitly mentioned and properly respected?
    \item[] Answer: \answerYes{}
    \item[] Justification: All evaluated LLMs and external benchmarks are cited (Section~\ref{sec:exp_setup}, Appendix~\ref{app:exp_details}) and used under their terms of use: commercial APIs for the closed-weight models, the licenses on the HuggingFace model cards for the open-weight 32B models, and the licenses of the original publications for the benchmarks. Gurobi 11.0 is used under an academic license.

\item {\bf New assets}
    \item[] Question: Are new assets introduced in the paper well documented and is the documentation provided alongside the assets?
    \item[] Answer: \answerYes{}
    \item[] Justification: RetailOpt-190 is documented in Section~\ref{sec:benchmark} and Appendices~\ref{app:reference_milp}--\ref{app:data_format}: the 38 archetypes in 8 families (Appendix~\ref{app:families}), the reference MILP (Appendix~\ref{app:reference_milp}), the JSON schema and data access patterns (Appendix~\ref{app:data_format}), the instance generation procedure (Appendix~\ref{app:perturbation}), and the ground-truth solver settings (Appendix~\ref{app:solver_config}). It is released with the code at the repository listed above.

\item {\bf Crowdsourcing and research with human subjects}
    \item[] Question: For crowdsourcing experiments and research with human subjects, does the paper include the full text of instructions given to participants and screenshots, if applicable, as well as details about compensation (if any)? 
    \item[] Answer: \answerNA{}
    \item[] Justification: The research involves no crowdsourcing and no human subjects.

\item {\bf Institutional review board (IRB) approvals or equivalent for research with human subjects}
    \item[] Question: Does the paper describe potential risks incurred by study participants, whether such risks were disclosed to the subjects, and whether Institutional Review Board (IRB) approvals (or an equivalent approval/review based on the requirements of your country or institution) were obtained?
    \item[] Answer: \answerNA{}
    \item[] Justification: No human subjects research is conducted, so IRB approval is not applicable.

\item {\bf Declaration of LLM usage}
    \item[] Question: Does the paper describe the usage of LLMs if it is an important, original, or non-standard component of the core methods in this research? Note that if the LLM is used only for writing, editing, or formatting purposes and does \emph{not} impact the core methodology, scientific rigor, or originality of the research, declaration is not required.
    \item[] Answer: \answerYes{}
    \item[] Justification: LLMs are the objects of evaluation (five models generate optimization code) and a component of the method: the L2 layer uses an LLM to extract candidate constraints and objective terms for solver-based perturbation testing (Section~\ref{sec:l2}, Appendix~\ref{app:reloop_details}). All configurations, prompts, and decoding parameters are documented in Appendices~\ref{app:data_format}--\ref{app:exp_details}.

\end{enumerate}

\end{document}